\documentclass[psfig]{aa}
\input psfig.tex

\begin{document}                                                                

\twocolumn


   \title{Detection of X-ray Clusters of Galaxies by Matching 
          RASS Photons and SDSS Galaxies within GAVO}

   \titlerunning{Detection of X-ray Clusters}

   \author{Peter\,\,Schuecker, Hans\,\,B\"ohringer and Wolfgang\,Voges}

   \authorrunning{Schuecker et al.}

   \offprints{Peter Schuecker\\ peters@mpe.mpg.de}

   \institute{
    Max-Planck-Institut f\"ur extraterrestrische Physik,
             Giessenbachstra{\ss}e 1, 85741 Garching, Germany\\ 
}
                                                                                
   \date{Received  ; accepted }                         
   
   \markboth{Detection of X-ray Clusters}{}

\abstract{A new method for a simultaneous search for clusters of
galaxies in X-ray photon maps and optical galaxy maps is
described. The merging of X-ray and optical data improves the source
identification so that a large amount of telescope time for
spectroscopic follow-up can be saved. The method appears thus ideally
suited for the analysis of the recently proposed wide-angle X-ray
missions like DUO and ROSITA.  As a first application, clusters are
extracted from the 3rd version of the ROSAT All-Sky Survey and the
Early Date Release of the Sloan Digital Sky Survey (SDSS).  The
time-consuming computations are performed within the German
Astrophysical Virtual Observatory (GAVO). On a test area of 140 square
degrees, 75 X-ray clusters are detected down to an X-ray flux limit of
$3-5\times 10^{-13}\,{\rm erg}\,{\rm s}^{-1}\,{\rm cm}^{-2}$ in the
ROSAT energy band 0.1-2.4\,keV. The clusters have redshifts $z\le
0.5$. The survey thus fills the gap between traditional large-area
X-ray surveys and serendipitous X-ray cluster searches based on
pointed observations, and has the potential to yield about 4,000 X-ray
clusters after completion of SDSS.  }

\maketitle

\section{Introduction}\label{INTRO}

Clusters of galaxies trace the peaks of the matter distribution and
their dynamics on cosmic scales (see review in Borgani \& Guzzo
2001). The corresponding cluster number density fluctuations can be
measured up to Gigaparsec scales where they display a simple Gaussian
random field with a high statistical significance (Schuecker et
al. 2002, 2003a). Therefore, the related mean spatial abundance, the
two-point correlation function, and the power spectrum give a useful
summary of these fluctuations. They are well fit by a flat, low
density Cold Dark Matter model without any significant contribution of
dark energy others than the cosmological constant (Schuecker et
al. 2003b). The data further support a simple scale-invariant biasing
scheme although the details are still discussed controversal (e.g.,
Bahcall 1988, Collins et al. 2000, Schuecker et al. 2001, Zandivarez,
Abadi \& Lambas 2001).

The usuability of rich clusters for cosmological investigations is
mainly related to their simple statistical properties and to their
comparatively simple physical structure (e.g. Kaiser 1986, Sarazin
1988, Neumann \& Arnaud 2001).  This is especially the case for X-ray
clusters of galaxies where the X-ray flux limit provides a clean
measure of the cluster selection function. A crucial point is the
reliability of cluster mass estimates (Schindler \& M\"uller 1993,
Evrard, Metzler \& Navarro 1996) where evolutionary effects could
further complicate the tests (Vikhlinin et al. 2002). However, for not
too distant systems the effects are found to be small (e.g. Gioia et
al. 2001, Rosati, Borgani \& Norman 2001) so that a proper calibration
with detailed X-ray observations appears to be possible and is already
in progress (H. B\"ohringer et al., in preparation).

Of fundamental importance for every precise cosmological test is the
construction of large, homogeneously selected and thus representative
samples.  Unfortunately, the time-consuming spectroscopic follow-up
observations of the cluster candidates selected from X-ray data
hampered the construction of X-ray cluster samples with sizes of the
order of $10^3$ or more (see review in Edge 2003). A similar problem
is expected for the recently proposed large-area X-ray missions like
DUO and ROSITA. From the construction of the Northern ROSAT All-Sky
(NORAS, B\"ohringer et al. 2000) and ROSAT ESO Flux-Limited X-ray
(REFLEX, B\"ohringer et al. 2003) samples we learned, however, that a
careful pre-selection of cluster candidates can lead to high success
rates during the spectroscopic follow-up observations and can thus
save a large amount of telescope time (see also B\"ohringer et
al. 1998, 2001, Guzzo et al. 1999).

In the following section we give a description of a general likelihood
method for the detection of clusters of galaxies (Sect.\,\ref{DETECT})
which allows a simultaneous analysis of multiwavelength data. In
Sect.\,\ref{RASS} we apply the method to the X-ray photon
distributions as given by the 3rd version of the ROSAT All-Sky Survey
(RASS-3, Voges et al. 1999). A new source detection in RASS is
necessary because it will turn out that -- because we combine X-ray
data with much deeper optical data -- the new method can go
significantly deeper compared to the standard analyses of RASS. For
the measurement of the source extent, hardness ratio and other
important X-ray quantities the Growth Curve Analyses (GCA) method of
B\"ohringer et al. (2000, 2001) is used. In Sect.\,\ref{SDSS} we apply
the same likelihood method to the galaxy distribution as given by the
Early Date Release of the Sloan Digital Sky Survey (SDSS, York et
al. 2000) using the $r'$ band where completeness and sample sizes are
highest down to the faintest magnitudes and which serves as a
`reference' passband where most methods using SDSS data are first
illustrated (future applications will, however, include all five SDSS
bands).  The similarity of the source detection in X-rays and cluster
search in the optical allows a straightforward merging of the two
results (Sect.\,\ref{RS}).

In a test area of 140 square degrees, 75 clusters are detected down to
the nominal flux-limit of $3-5\times 10^{-13}\,{\rm erg}\,{\rm
s}^{-1}\,{\rm cm}^{-2}$ in the ROSAT energy band $0.1-2.4$\,keV, that
is, 5--10 times deeper than REFLEX.  The measured parameters of the
clusters are described in Sect.\,\ref{CAT}. Extrapolated to a final
SDSS area of 7,000 square degrees, our method is thus expected to
yield about 4,000 X-ray clusters. To test the reliability of our
method, we compare in Sect.\,\ref{DISCUSS} our X-ray cluster sample
with various SDSS and RASS-based catalogues. Further tests with
realistic simulations are postponed to a second paper.

It will be seen that a detailed analysis of the RASS and SDSS data
requires massive computing power. Moreover, our method allows the
inclusion of further high-resolution maps from, e.g., temperature
anisotropies of the cosmic microwave background radiation, centimeter
or millimeter (Sunyaev-Zel'dovich) maps etc. The combined analysis of
such large databases is a typical example where virtual observatories
can help with a fast and easy data handling and a distribution of the
time-consuming computations over many computer systems (grid
computing). The present investigation was thus planned already from
its beginning to be performed within the German Astrophysical Virtual
Observatory (GAVO).

All computations assume a pressureless Friedmann-Le\-ma\^{\i}\-tre
world model with the Hubble constant $H_0$ in units of
$h=H_0/(100\,{\rm km}\,{\rm s}^{-1}\,{\rm Mpc}^{-1})$, the normalized
cosmological constant $\Omega_\Lambda=0.73$, and the density parameter
$\Omega_m=0.27$, but the resulting cluster catalogues do not depend on
the chosen values of the cosmic energy densities.

\section{A general method for cluster detection}\label{DETECT}

In the following we want to define likelihood functions appropriate
for the detection of clusters in point patterns. For the analysis a
quantity $L_V(\vec{x}_1,$ $\ldots,\vec{x}_N)$ $d\vec{x}_1\cdots
d\vec{x}_N$ is defined which gives the probability of finding exactly
one point in each of the infinitesimal volume elements
$d\vec{x}_1,\ldots,d\vec{x}_N$ centered at
$\vec{x}_1,\ldots,\vec{x}_N$. We assume that this probability is
Poissonian with the intensity parameter $\lambda$. The presence of a
cluster suggests that $\lambda$ is closely related to the projected
cluster profile (plus background) and is thus a function of
$\vec{x}$. The probability to find a point in a given volume element
is thus guided by $\lambda(\vec{x})$, and is independent of the
probability to find somewhere else another point. Because of this
independence property, $L_V$ is simply the product of $N$ terms, each
of the form $\lambda(\vec{x}_i)d\vec{x}_i$.

However, the test for the presence of a cluster is still incomplete
because we did not include the fact that no further points were
observed outside the $N$ infinitesimal volume elements. This
probability is given by the void probability function
$\exp(-\int\lambda(\vec{x})d\vec{x})$ where the integration extends
over the volume $V$ reduced by the sum of the infinitesimal volume
elements, $d\vec{x}_i$, which can be neglected due to their small size
so that, in total, the integration extends over the complete volume
$V$. Therefore, an appropriate likelihood function for cluster search
is of the form
\begin{equation}\label{L1}
\ln
L_V\,=\,-\int_V\lambda(\vec{x})d\vec{x}\,+\,\sum_{i=1}^N\,
\ln\lambda(\vec{x}_i)\,,
\end{equation}
which assumes high values when the model distribution
$\lambda(\vec{x})$ fits the observed pattern of $N$ points. Often,
additional properties of each point, like its magnitude, colour or
energy are available. This information can be directly incorporated,
extending the dimensionality of (\ref{L1}) by a further space which
includes the additional marks $m$,
\begin{eqnarray}\label{L5}
\ln L_{V\times M}\,&=&\,-\int_V\int_M\,\lambda(\vec{x},m)
dm\,d\vec{x}\,+\, \sum_{i=1}^N\,\ln\lambda(\vec{x}_i,m_i)\,.\nonumber\\ &
&
\end{eqnarray}
A mathematical exact derivation of (\ref{L5}) is given in Schuecker \&
B\"ohringer (1998) where the correctness of this model was verified
with a large number of Monte-Carlo experiments. Such additional tests
are necessary because sometimes likelihood estimates can yield biased
results. Different versions of this matched filter can be found in
e.g. Postman et al. (1996), Kepner et al. (1999), Kim et
al. (2002). The corresponding detection probability is
\begin{equation}\label{PROB}
P\,\equiv\,\frac{L-L_0}{L}\,=\,1\,-\,
\exp{\left(-\ln\frac{L}{L_0}\right)}\,,
\end{equation}
with the likelihood ratio $L/L_0$ giving a measure of the probability
for the presence of a cluster normalized to the probability for the
absence of a cluster (blank field).  In the following sections we will
specialize Eq.\,(\ref{L1}) to the analysis of the spatial distribution
of RASS photons, and Eq.\,(\ref{L5}) to the analysis of the spatial
and magnitude distributions of SDSS galaxies. The likelihood for a
cluster present in both X-rays and optical is then simply given by the
point-wise product of the two likelihood maps.

\section{Cluster detection in RASS-3}\label{RASS}

For the cluster detection in RASS-3 we assume $\lambda$ to be
proportional to the {\it local} countrate $c(\vec{\theta})$, giving
the number of X-ray photon events observed in a unit time interval in
a specific energy band per arcsec$^2$ at the angular coordinate
$\vec{\theta}$ measured (in arcsec) from the centre of the region to
be tested for the presence of a cluster. If $t(\vec{\theta})$ is the
exposure time (in seconds) of RASS-3 at $\vec{\theta}$ then
Eq.\,(\ref{L1}) translates into
\begin{equation}\label{R1}
\ln L\,=\,-\int_0^\infty
d^2\vec{\theta}\,c(\vec{\theta})\,t(\vec{\theta})\,+\,
\sum_{i=1}^N\,\ln\left[c(\vec{\theta_i})\,t(\vec{\theta_i})\right]\,.
\end{equation}
For the local countrate we assume the model
$c(\vec{\theta})=\bar{c}+\Lambda\,\tilde{P}(\vec{\theta}/\theta_c)$: a
local background countrate $\bar{c}$ plus a cluster with the
(convolved) apparent cluster profile
$\tilde{P}(\vec{\theta}/\theta_c)$ and the projected core radius
$\theta_{\rm c}$. With the normalization
\begin{equation}\label{R5}
\int_0^\infty\,d^2\vec{\theta}\,\tilde{P}(\vec{\theta}/\theta_c)\,=\,1\,,
\end{equation}
and the definition of an effective exposure time
\begin{eqnarray}\label{R10}
t_{\rm eff}\,=\,\int_0^\infty\,d^2\vec{\theta}\,
\tilde{P}(\vec{\theta}/\theta_c)\,t(\vec{\theta})\,,
\end{eqnarray}
we get the likelihood for a cluster X-ray source in units of the
likelihood for the local background without any source,
\begin{eqnarray}\label{R15}
\ln\left(\frac{L}{L_0}\right)\,=\,-\int_0^\infty
d^2\vec{\theta}\,\Lambda\,\tilde{P}(\vec{\theta}/
\theta_c)\,t(\vec{\theta})\,+\nonumber\\
\sum_{i=1}^N\ln\frac{\bar{c}\,+\,\Lambda\,\tilde{P}(\vec{\theta}_i/
\theta_c)}{\bar{c}}\,,
\end{eqnarray}
where $L_0$ is given by (\ref{R1}) for $\Lambda=0$. From (\ref{R15})
and the maximization condition $\partial\ln(L/L_0)/\partial\Lambda=0$
the following two equations are obtained,
\begin{equation}\label{R20}
t_{\rm eff}\,=\,\sum_{i=1}^N\frac{\tilde{P}(\vec{\theta}_i/\theta_c)}
{\bar{c}+\Lambda\,\tilde{P}(\vec{\theta}_i/\theta_c)}\,,
\end{equation}
\begin{equation}\label{R25}
\ln\left(\frac{L}{L_0}\right)\,=\,-\Lambda\,t_{\rm eff}\,+\,
\sum_{i=1}^N
\ln\frac{\bar{c}+\Lambda\,\tilde{P}(\vec{\theta}_i/\theta_c)}{\bar{c}}\,,
\end{equation}
which are used for cluster detection and countrate determination in
the following manner: For a chosen Right Ascension, Declination and
redshift, the cluster profile in form of a standard $\beta$ profile
(Cavaliere \& Fusco-Femiano 1976) with the core radius $r_c$ and the
slope parameter $\beta$ is transformed from metric to angular scales
$\theta_c$ and convolved with the average RASS-3 pointspread function
of the X-ray energy band to give $\tilde{P}(\vec{\theta}/
\theta_c)$. The exposure times $t(\vec{\theta})$ are computed for each
position covered by RASS-3 and yield the effective exposure time from
Eq.\,(\ref{R10}), the countrate from (\ref{R20}), and thus the
likelihood ratio from (\ref{R25}). The filter is shifted over the
three-dimensional survey volume so that sky coordinates, redshifts and
countrates which simultaneously maximize Eq.\,(\ref{R25}) indicate the
presence of an X-ray cluster with the detection probability
(Eq.\,\ref{PROB}).

\begin{figure}
\vspace{-1.3cm}
\centerline{\hspace{0.7cm}
\psfig{figure=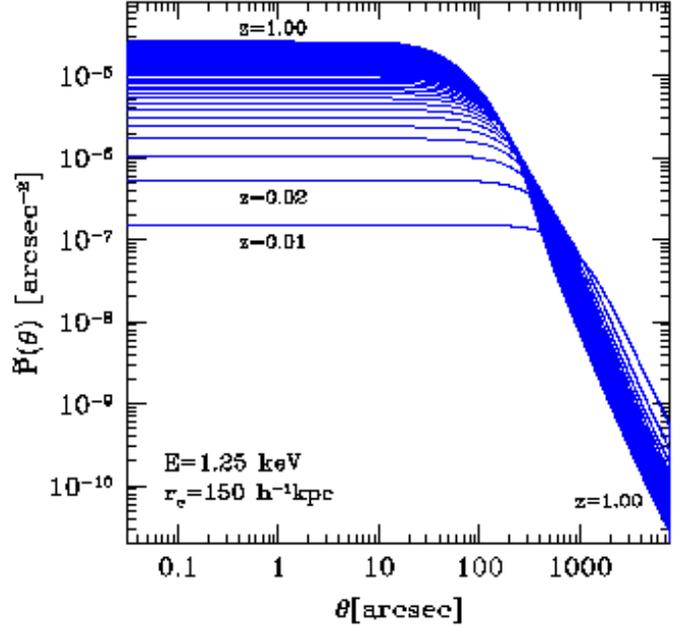,height=10.5cm,width=10.5cm}}
\vspace{-1.00cm}
\caption{\small X-ray $\beta$ profile of galaxy clusters with the
slope parameter $\beta=2/3$, the core radius $150\,h^{-1}\,{\rm kpc}$
at different redshifts, convolved with the point spread functions of
RASS-3 for the energy $1.25$\,keV and normalized according to
Eq.\,\ref{R5}. The RASS-3 point spread function closely resembles the
curve labeled $z=1.00$.}
\label{FIG_PSF}
\end{figure}

\begin{figure*}
\vspace{-0.0cm}
\centerline{\hspace{0.0cm}
\psfig{figure=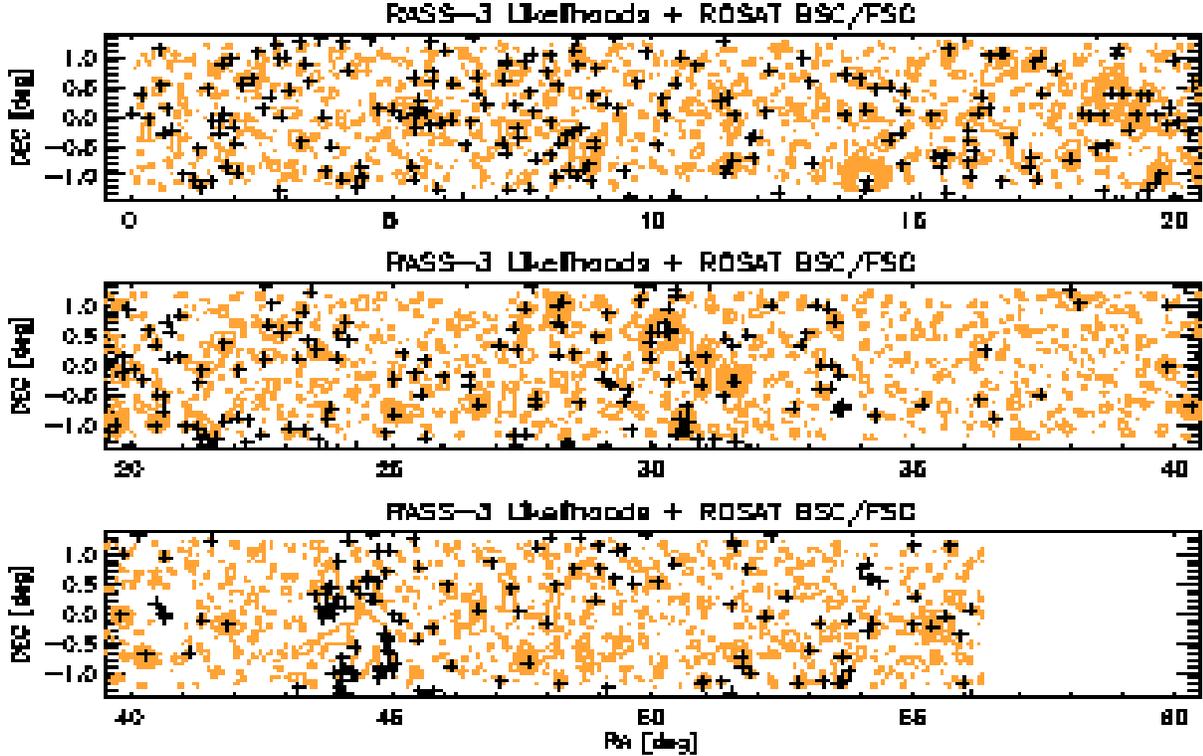,height=11.0cm,width=17.0cm}}
\vspace{-0.50cm}
\caption{\small Maximum likelihood contours ($1\sigma$, $2\sigma$ ...)
for RASS-3 in the energy range 0.5-2.0\,keV. Crosses mark objects
found in the merged ROSAT Bright and Faint (RBF) Source
Catalogues. Bright X-ray sources at $({\rm R.A.}=20.7^\circ,{\rm
DEC}=-0.70^\circ)$, $({\rm R.A.}=33.65^\circ,{\rm DEC}=-0.775^\circ)$,
$({\rm R.A.}=40.675^\circ,{\rm DEC}=-0.025^\circ)$ and $({\rm
R.A.}=54.20^\circ,{\rm DEC}=0.60^\circ)$ have been removed from the
RASS-3 photon maps. Note that RBF sources are also shown slightly
outside of the SDSS Declination range $[-1.25,+1.25]$\,degrees.}
\label{FIG_LOG0}
\end{figure*}

\begin{figure*}
\vspace{-0.0cm}
\centerline{\hspace{0.0cm}
\psfig{figure=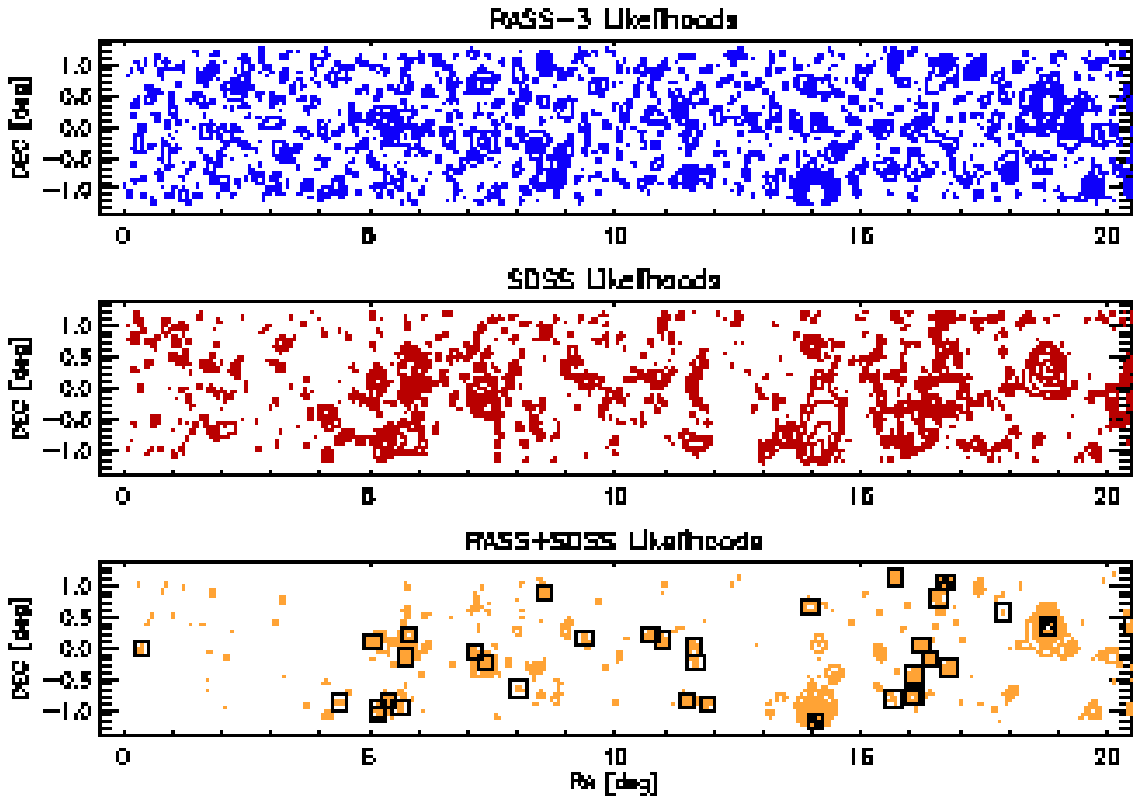,height=11.0cm,width=17.0cm}}
\vspace{+1.0cm}
\centerline{\hspace{0.0cm}
\psfig{figure=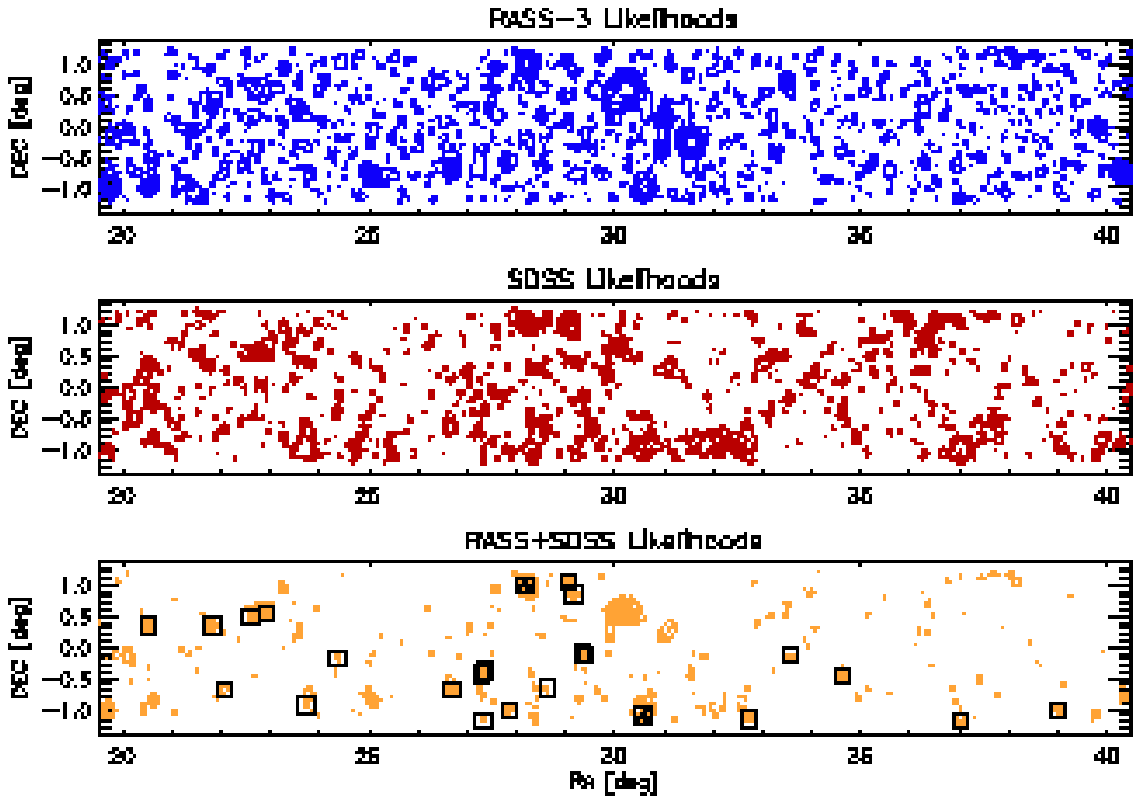,height=11.0cm,width=17.0cm}}
\vspace{-0.50cm}
\caption{\small Maximum likelihood contours based on RASS-3 X-ray
photons (1st and 4th panels, $1\sigma$, $2\sigma$ ... contours), SDSS
galaxies (2nd and 5th panels, $\ge 10\sigma$) and the combined maximum
likelihood contours of RASS-3 and SDSS data (3rd and 6th panels, $\ge
10\sigma$). Crosses mark the position of the deepest X-ray cluster
samples available sofar (REFLEX-2, X-ray flux limit $1.8\times
10^{-12}\,{\rm erg}\,{\rm s}^{-1}\,{\rm cm}^{-2}$). Squares mark the
position of the X-ray clusters of the final sample. The structure at
${\rm R.A.}=30$\,deg and ${\rm DEC}=+0.5$\,deg has a too low
likelihood in the SDSS galaxy distribution and is thus not included in
the final catalogue.}
\label{FIG_LOG1}
\end{figure*}

\begin{figure*}
\vspace{-0.0cm}
\centerline{\hspace{0.0cm}
\psfig{figure=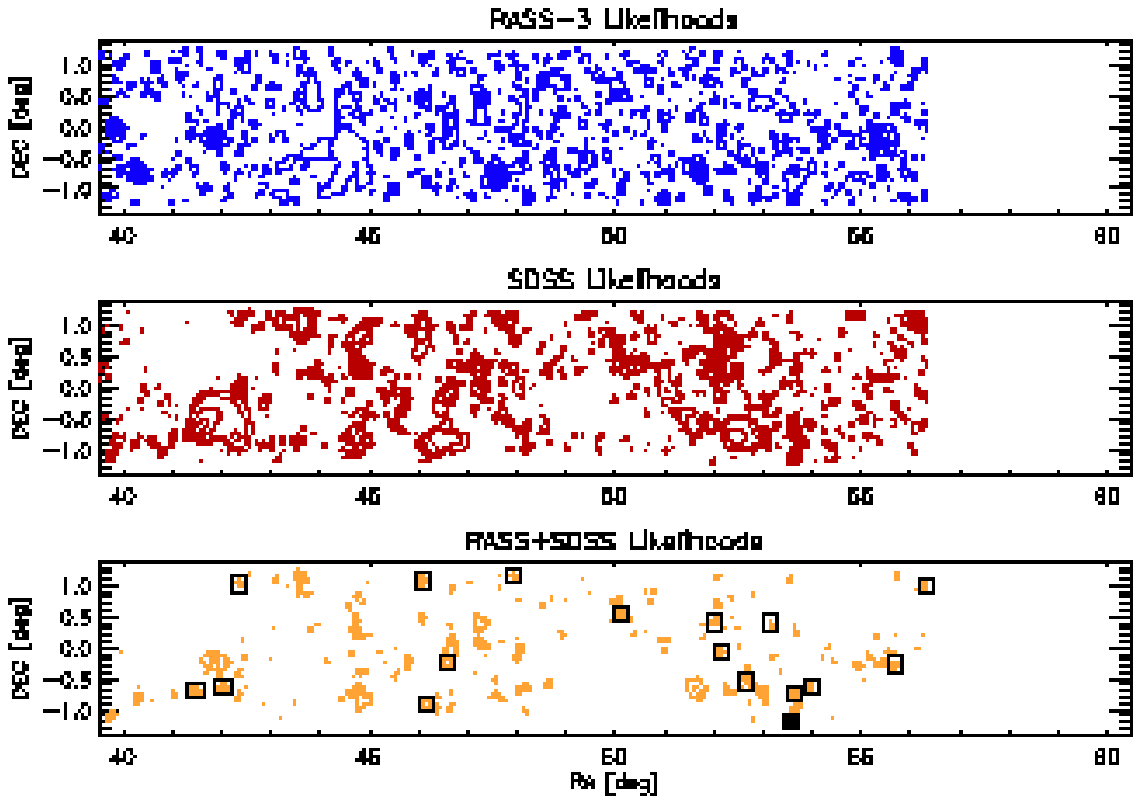,height=11.0cm,width=17.0cm}}
\vspace{-0.50cm}
\caption{\small Maximum likelihood contours. Symbols as in
Fig.\,\ref{FIG_LOG1}. Squares mark the position of the X-ray clusters
of the final sample.}
\label{FIG_LOG2}
\end{figure*}

Distant and/or compact X-ray clusters might appear point-like in
RASS-3. For such sources the countrates converge to quite stable
values when the cluster filter slides towards higher redshifts. The
same is true for real point-like X-ray sources. Therefore, the
cluster-finder described above can also be used for the detection of
point sources which can be identified by their high value of the
estimated redshift (usually the maximum $z$ of the tested redshift
range). 

Another important quantity is the number of source counts, i.e., the
number of X-ray photons above the X-ray background:
\begin{equation}\label{NPH}
N_{\rm ph}\,=\,\Lambda\,t_{\rm eff}\,.
\end{equation}
The number of source photons gives a further handle to evaluate the
significance of the detected X-ray source and the expected errors of
X-ray countrate and fluxes plus an additional error of the background.

\subsection{Application to RASS-3}\label{APPL1}

For the cluster detection on RASS-3 we choose a test area of 140
square degrees with $0.0\le{\rm R.A.}\le 56$ deg and $-1.25\le{\rm
DEC.}\le +1.25$ deg (part of the SDSS Early Data Release). Our
analysis is restricted to the ROSAT energy range 0.5-2.0\,keV where
the contrast between X-ray background and cluster is highest
(B\"ohringer et al. 2000). GAVO provides {\it one} photon event file
and {\it one} exposure map of the complete sky based on RASS-3. In the
test area and the selected energy band we extracted 180,466 X-ray
photons. The median exposure time is 369 seconds which is close to the
median exposure time of 388 seconds of the total RASS-3.  We assume a
standard $\beta$ profile with a fixed shape parameter of $\beta=2/3$
and a fixed core radius of $r_{\rm c}=150\,h^{-1}$\,kpc.  The profiles
are convolved with the spatially averaged point spread function of
RASS-3 at 1.25\,keV (Boese 2000). A sample of normalized cluster X-ray
profiles (see Eq.\,\ref{R5}) is plotted in Fig.\,\ref{FIG_PSF}.

For the matched filter only the photons located in a quadratic
detection cell with a total area of 1.0 square degree are used
(typically 1,200 photons in the energy band 0.5-2.0\,keV). This
detection cell is shifted by increments of 1.5 arcmin along R.A. and
DEC. over the test area. The size of the increment is comparable to
the half power radius (96 arcsec) of the point spread function of
RASS-3 at 1\,keV. In redshift direction the increment is $\Delta
z=0.02$ in the range $0.02\le z \le 1.0$, consistent with the error
estimates obtained with the optical data (Sect.\,\ref{RS}). Future
investigations will have more computing power and will decrease the
increment sizes in all three directions significantly. The local
background countrate is estimated with formal Poisson errors $<3$\% by
the median of the exposure-weighted countrates obtained in a large
number of subcells per detection cell. Small background errors are of
fundamental importance for the detection of very faint X-ray sources.

The comparatively large size of the detection cell yields stable
background estimates. However, bright point sources with countrates of
more than about $0.5\,{\rm s}^{-1}$ affect the source detection and
background determination significantly. They are detected in a first
run of the matched filter and then removed from the RASS-3 photon
event file. In the present case we deleted the photons of on average
one bright point source per 40 square degrees. After their removal, an
almost unbiased search for faint sources can be performed in a second
run (see Figs.\,\ref{FIG_LOG0}-\ref{FIG_LOG2}).

Figure\,\ref{FIG_LOG0} compares the likelihood contours as obtained
with the matched filter and the source positions as published in the
combined ROSAT Bright and Faint (RBF, Voges et al. 1999) Source
Catalogues (crosses). The redshift direction is supressed by replacing
the likelihood values obtained along the $z$ direction by their
maximum $\ln(L/L_0)$ value. The latter likelihood together with the
R.A. and DEC. are shown. Not shown is the redshift where the maximum
likelihood is obtained and the corresponding countrate. In the
following we will call these two-dimensional $\ln(L/L_0)$ arrays {\it
maximum} likelihood maps. The maximum likelihood contours start at
$\ln(L/L_0)=1.0$ which corresponds to a detection probability of
$P=0.63$, i.e., at the $\sim1\sigma$ detection limit of RASS-3.

A detailed comparison of the results in Sect.\,\ref{DISCUSS} shows
that the matched filter finds more X-ray clusters than published in
RBF. This is due to the fact that the matched filter uses both the
RASS and the SDSS data which helps to go below the usual detection
threshold of the RBF. We also have a preliminary MPE-internal list of
RASS-3 sources obtained with the same detection methods as used for
the extraction of the RBF sources on RASS-2, but without a detailed
visual screening of the detections.  The quantitative comparison in
Sect.\,\ref{DISCUSS} includes this preliminary source list.

Figures\,\ref{FIG_LOG1} and \ref{FIG_LOG2} show (again) the maximum
likelihood contours obtained with the RASS-3 survey in the test area,
but now in comparison to maximum likelihood maps obtained with the
SDSS data and maximum likelihood maps obtained from the combination of
RASS and SDSS data. The determination of the SDSS maximum likelihood
maps is described in Sect.\,\ref{SDSS}, the combination of RASS and
SDSS maximum likelihood maps and preparation of the final cluster
sample in Sect.\,\ref{RS}. REFLEX-2 clusters and the X-ray clusters of
our final RASS/SDSS sample are shown as crosses and squares,
respectively, on the combined maps. The REFLEX-2 clusters are clearly
visible as the most significant clusters and illustrate that the
present method reaches much deeper X-ray flux limits compared to
existing RASS-based cluster catalogues.

\section{Cluster detection in SDSS}\label{SDSS}

The detection of clusters in SDSS is based on Eq.\,(\ref{L5}). Our
analytic treatment follows Sect.\,\ref{RASS} and is described in
detail in Schuecker \& B\"ohringer (1998). The final equations are
\begin{equation}\label{S20}
1\,=\,\sum_{i=1}^N\frac{P(\vec{\theta}_i/\theta_c)\,\phi(m_i-m^*)}
{\bar{c}+\Lambda\,P(\vec{\theta}_i/\theta_c)\,\phi(m_i-m^*)}\,,
\end{equation}
\begin{equation}\label{S25}
\ln\left(\frac{L}{L_0}\right)=-\Lambda+\sum_{i=1}^N
\ln\left[1+\frac{\Lambda\,P(\vec{\theta}_i/\theta_c)\,
\phi(m_i-m^*)}{b(m_i)}\right]\,,
\end{equation}
which correspond to the Eqs.\,(\ref{R20},\ref{R25}) derived for the
X-ray data. In Eqs.\,(\ref{S20}, \ref{S25}), $\Lambda$ is now the
optical richness which is corrected as described in Schuecker \&
B\"ohringer (1998), $b(m)$ the average number counts of `background'
galaxies with apparent magnitude $m$, $P(\cdot)$ a King-like profile,
and $\phi(\cdot)$ the apparent Schechter luminosity function with
characteristic apparent magnitude $m^*$. Compared to the optical
richness, we regard the X-ray luminosity as more directly related to
the total gravitating cluster mass (see, e.g., Borgani \& Guzzo
2001). Therefore, in the following we do not much concentrate on a
discussion of the optical richnesses.

\subsection{Application to SDSS}\label{APPL2}

For the optical cluster search, SDSS galaxies with
extinction-corrected {\it model} magnitudes $r'\le 21.0$\,mag are
used. Above this magnitude limit a robust star/galaxy separation is
expected (Scranton et al. 2003). In our test area we have 2,086,654
SDSS galaxies. A detailed description of our galaxy selection,
especially the choice of SDSS catalogue flags, is given in Sect.\,4 of
Popesso et al. (2003).

For the optical cluster search a truncated King profile with a core
radius of $150\,h^{-1}$\,kpc and a cutoff radius of $1\,h^{-1}$\,Mpc
is used. The magnitude distribution of the background galaxies is
obtained in detection cells with an area of $5.0\times 2.5$ square
degrees (about 180,000 galaxies per detection cell). Although this
area might be regarded as comparatively large for the determination of
the {\it local} galaxy background, we prefer to work with a larger
detection area because the RASS-3 clusters are mainly at comparatively
small redshifts around $z\approx 0.15$ (Sect.\,\ref{DISCUSS}) where a
good measurement of the bright end of the galaxy luminosity function
is needed. A larger area is also better for the detection of nearby
less rich systems (Sect.\,\ref{DISCUSS}). 

For the cluster luminosity function we assume a Schechter function
with the characteristic magnitude $M^*_{r'}=-21.43+5\log h$ and the
faint end slope of $\alpha=-0.90$ as obtained from the composite
luminosity function of SDSS clusters (Goto et al. 2002b). For the
conversion of absolute into apparent magnitudes we assumed the
$K$-corrections for the early-type galaxies as given in Fukugita,
Shimasaku \& Ichikawa (1995). The richness estimates are obtained
3.0\,mag below the apparent Schechter characteristic magnitude.

In order to allow a direct combination of the likelihood contours, the
values of the increment parameters in Right Ascension, Declination,
and redshift used for the analysis of the SDSS data are the same as
for the X-ray data. The maximum likelihood contours are shown in
Figs.\,\ref{FIG_LOG1}-\ref{FIG_LOG2}.

\section{Combined cluster detection in RASS-3 and SDSS}\label{RS}

The similarity of the likelihood filters applied in X-rays and the
optical, and the identical sampling of the three-dimensional
likelihood parameter space (R.A., DEC., $z$) makes the combination of
the results obtained with both data sets by point-wise multiplication
of the likelihood values in principle straightforward. However, it
turned out that the cluster redshift estimates obtained from the SDSS
data are much better compared to the estimates obtained from the RASS
data (see Tabs.\,2-3). This is mainly caused by the limited angular
resolution of RASS. Therefore, we define the final angular coordinates
R.A. and DEC. of a cluster by the location of a peak in the multiplied
RASS and SDSS maximum likelihood map
(Figs.\,\ref{FIG_LOG1}-\ref{FIG_LOG2}). The redshift of a cluster is
defined by the location of a peak along the $z$ direction in the
likelihood map of the SDSS data at the final (R.A.,DEC.) coordinates.

The naive multiplication of maximum likelihood maps from X-ray and
optical data and the subsequent peak analyses in the combined map can
lead to the detection of spurious X-ray clusters when a strong signal
from one energy band is combined with a spurious signal from the other
band.  The resulting systematics can be reduced by analysing only
those peaks in the combined likelihood maps where the related peaks in
the X-ray and in the optical maps exceed certain likelihood
thresholds.

\begin{figure}
\vspace{-2.2cm}
\centerline{\hspace{1.2cm}
\psfig{figure=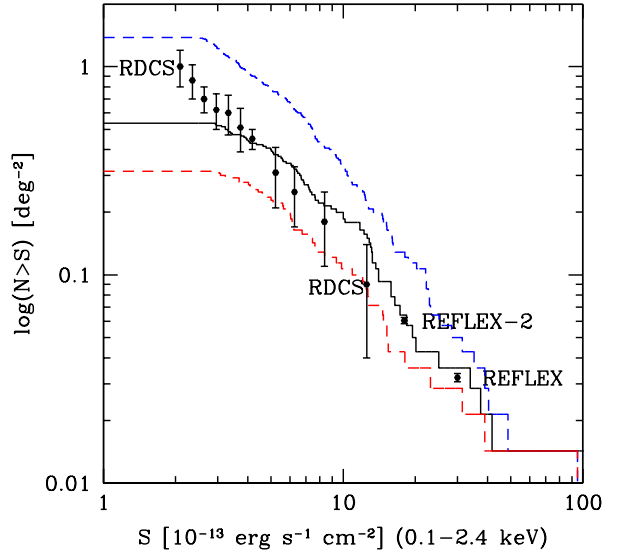,height=11.0cm,width=11.0cm}}
\vspace{-1.1cm}
\caption{\small Cumulative X-ray cluster number counts of the
RASS/SDSS clusters (histograms) for a log-likelihood minimum of 15
applied to the SDSS data (continuous line), for 25 (lower dashed
line), and for 5 (upper dashed line). The RASS/SDSS cluster counts are
compared with results obtained with other surveys (points with
$1\sigma$ error bars: with RDCS from Rosati et al. 1998 converted to
the ROSAT energy band 0.1--2.4\,keV assuming a Raymond-Smith spectrum
with $z=0$, $k_{\rm B}T=5$\,keV, and solar abundance, with REFLEX, and
REFLEX-2). No corrections for variations of the angular survey
sensitivity (effective survey area) are applied to the RASS/SDSS and
REFLEX-2 data {\rm (see main text)}.}
\label{FIG_LOGN}
\end{figure}

\begin{figure}
\vspace{-2.2cm}
\centerline{\hspace{1.2cm}
\psfig{figure=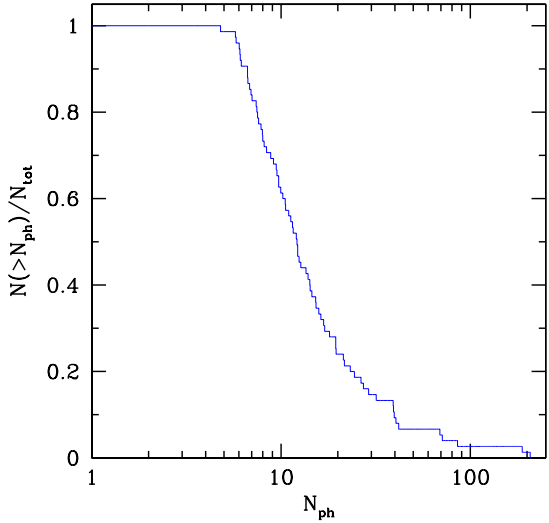,height=11.0cm,width=11.0cm}}
\vspace{-1.2cm}
\caption{\small Fraction of sources with minimum source counts $N_{\rm
ph}$ as obtained with the matched filter for the final X-ray cluster
sample derive in Sect.\,\ref{RS}.}
\label{FIG_NPH_HST}
\end{figure}

\begin{figure}
\vspace{-2.2cm}
\centerline{\hspace{1.2cm}
\psfig{figure=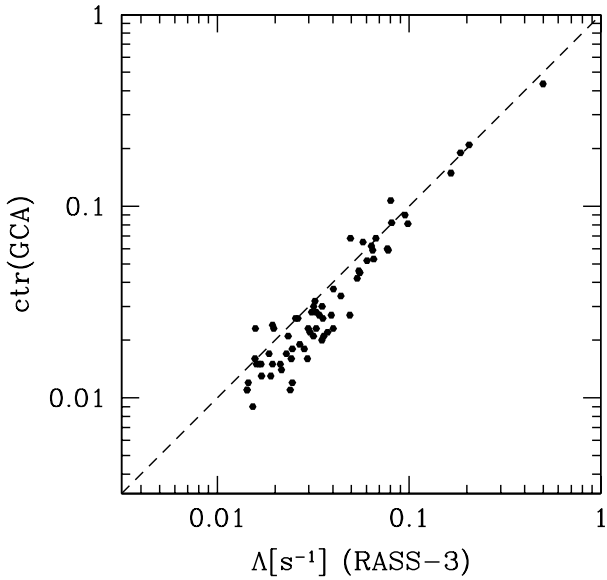,height=11.0cm,width=11.0cm}}
\vspace{-0.8cm}
\caption{\small Comparison of countrates obtained with the matched
filter $\Lambda$ and with the growth curve analysis ctr(GCA) for the
final X-ray cluster sample. Not shown are clusters partially outside
the test area (1 cluster) and clusters where GCA could not find a
clear flattening of the background-subtracted countrates beyond the
so-called outer significance radius (6 clusters).}
\label{FIG_GCA}
\end{figure}

\begin{figure}
\vspace{-2.2cm}
\centerline{\hspace{1.2cm}
\psfig{figure=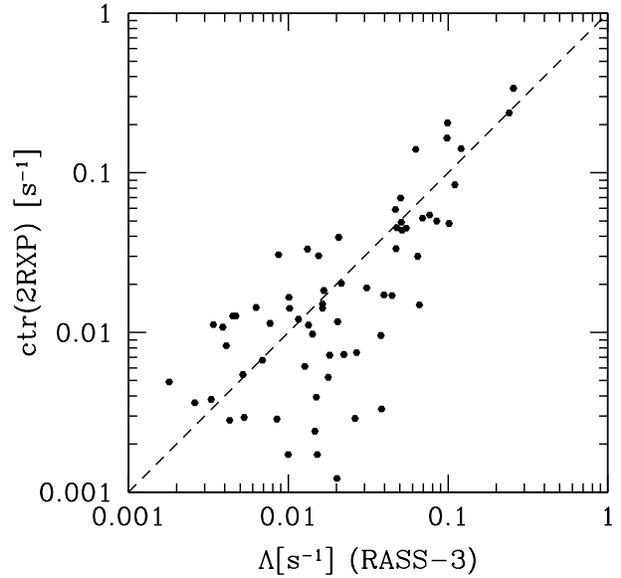,height=11.0cm,width=11.0cm}}
\vspace{-0.8cm}
\caption{\small Comparison of countrates obtained from RASS-3 with
countrates as given in the 2nd ROSAT Source Catalog of Pointed
Observations with the Position Sensitive Proportional Counter. The
comparison includes all types of X-ray sources.}
\label{FIG_CTR}
\end{figure}

\begin{figure}
\vspace{-2.5cm}
\centerline{\hspace{0.5cm}
\psfig{figure=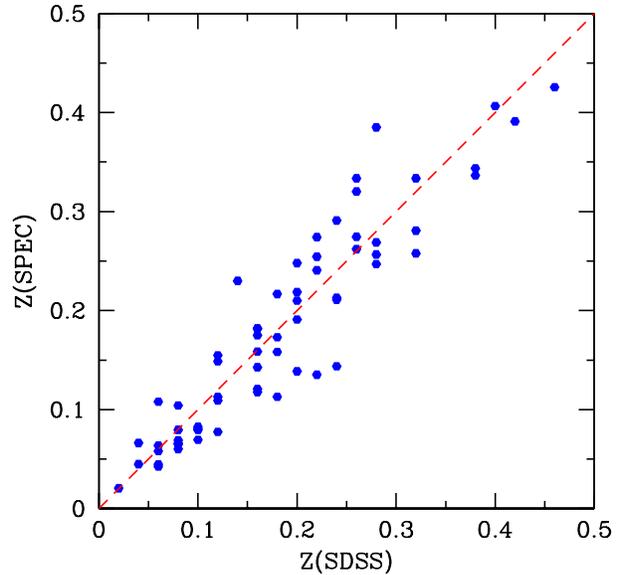,height=12.0cm,width=12.0cm}}
\vspace{-1.6cm}
\caption{\small Comparison of cluster redshifts estimated from the
SDSS data with spectroscopic cluster redshifts (see Tabs.\,2-3).}
\label{FIG_ZZ}
\end{figure}

The minimum likelihood threshold for the X-ray data is determined by
our goal to search for X-ray clusters down to the detection limit of
RASS. In the present investigation we thus use a minimum likelihood of
$\ln(L/L_0)=3.0$ for RASS-3 which corresponds to a detection
probability of $P=0.95$. The resulting error rate in RASS is
compensated by the high minimum likelihood threshold for SDSS of
$\ln(L/L_0)=15$. This value is found to approximately reproduce the
detection of all known comparatively bright REFLEX clusters and limits
the number of spurious detections as shown by the flux/number counts
in Fig.\,\ref{FIG_LOGN}. The determination of cluster X-ray fluxes and
related quantities is described in Sect.\,\ref{CAT}. For this
preliminary calibration of the likelihood threshold the RDCS sample
was not included because the counts have comparatively large
statistical errors in this range. We show the raw RASS/SDSS cluster
counts uncorrected for the effective survey area in order to
illustrate more directly the results of the cluster search. The final
calibration will be performed over a much larger sky area including a
proper weighting with the effective survey area.

\begin{table}
{\bf Tab.\,1.} Comparison of countrates $\Lambda$ obtained from RASS-3
and as given in the 2RXP catalogue. Col.\,1: Minimum countrate
(RASS-3). Col.\,2: Number of X-ray sources. Col.\,3: Average
difference $({\rm ctr(RASS3)}$ minus ${\rm ctr(2RXP)})$. Col.\,4:
Standard deviation of the countrate differences. See also
Fig.\,\ref{FIG_CTR}.\\
\vspace{-0.1cm}
\begin{center}
\begin{tabular}{cccc}
\hline
\hline
 $\Lambda>$ & $N$ & $\Delta(\Lambda)$&$\sigma_\Lambda$\\
\hline
$0.100$&$ 5$&$-0.0039$&$0.0515$\\
$0.080$&$ 8$&$-0.0197$&$0.0593$\\
$0.060$&$13$&$-0.0084$&$0.0558$\\
$0.040$&$21$&$-0.0037$&$0.0445$\\
$0.020$&$32$&$+0.0028$&$0.0378$\\
$0.010$&$50$&$+0.0028$&$0.0306$\\
$0.008$&$52$&$+0.0024$&$0.0302$\\
$0.006$&$55$&$+0.0020$&$0.0294$\\
$0.004$&$61$&$+0.0016$&$0.0280$\\
$0.002$&$65$&$+0.0012$&$0.0271$\\
$0.001$&$66$&$+0.0011$&$0.0269$\\
\hline
\hline
\end{tabular}
\end{center}
\end{table}

\section{Cluster characteristics}\label{CAT}

In our test area, 75 clusters fulfill the criteria derived in
Sect.\,\ref{RS} (see Tabs.\,2-3, squares in
Figs.\,\ref{FIG_LOG1}-\ref{FIG_LOG2}). Their positions, source counts,
effective exposure times, countrates etc. are fixed by their values at
the local maxima in the combined maximum likelihood distributions. The
X-ray source counts $N_{\rm ph}$ are then determined with
Eq.\,(\ref{NPH}). The distribution function of the number of source
counts $N_{\rm ph}$ obtained with the final X-ray cluster sample (see
Sect.\,\ref{RS}) is shown in Fig.\,\ref{FIG_NPH_HST}. The cumulative
distribution shows that about 2/3 of the X-ray clusters have at least
10 X-ray photons in the energy range 0.5-2.0\,keV above the local
background yielding formal errors of the countrates of about 30\%
(plus negligible errors of $<3\%$ from the background).

Fig.\,\ref{FIG_GCA} compares the countrates obtained with the matched
filter $\Lambda$ and with GCA for our final cluster sample. For
countrates $\Lambda>0.05$/s, no significant differences between
$\Lambda$ and GCA countrates are seen. For $\Lambda<0.05\,{\rm
s}^{-1}$, the matched filter gives on average 15\% higher countrates
compared to GCA. We attribute this systematic to the fact that the
matched filter integrates the source photons out to the edge of the
detection cell (formally the integration extends to infinity) whereas
the GCA integrates the source photons out to the so-called outer
significance radius of a source. Therefore, the former method has the
danger of slightly overestimating the countrate, while the latter
method does not count the outer wings of the sources and thus slightly
underestimates the countrate. The differences between these two
integration schemes increase with decreasing countrate as seen in
Fig.\,\ref{FIG_GCA}.

GCA further gives for each cluster an additional estimate of the
missing flux in order to transform aperature to total X-ray
countrates. For example, at a flux of $3\times 10^{-12}\,{\rm
erg}\,{\rm s}^{-1}\,{\rm cm}^{-1}$ the average of the latter
correction is about 10\%. For the matched filter numerical simulations
are in preparation to measure the appropriate corrections.

\begin{figure*}
\vspace{0.0cm}
\centerline{\hspace{-9.0cm}
\psfig{figure=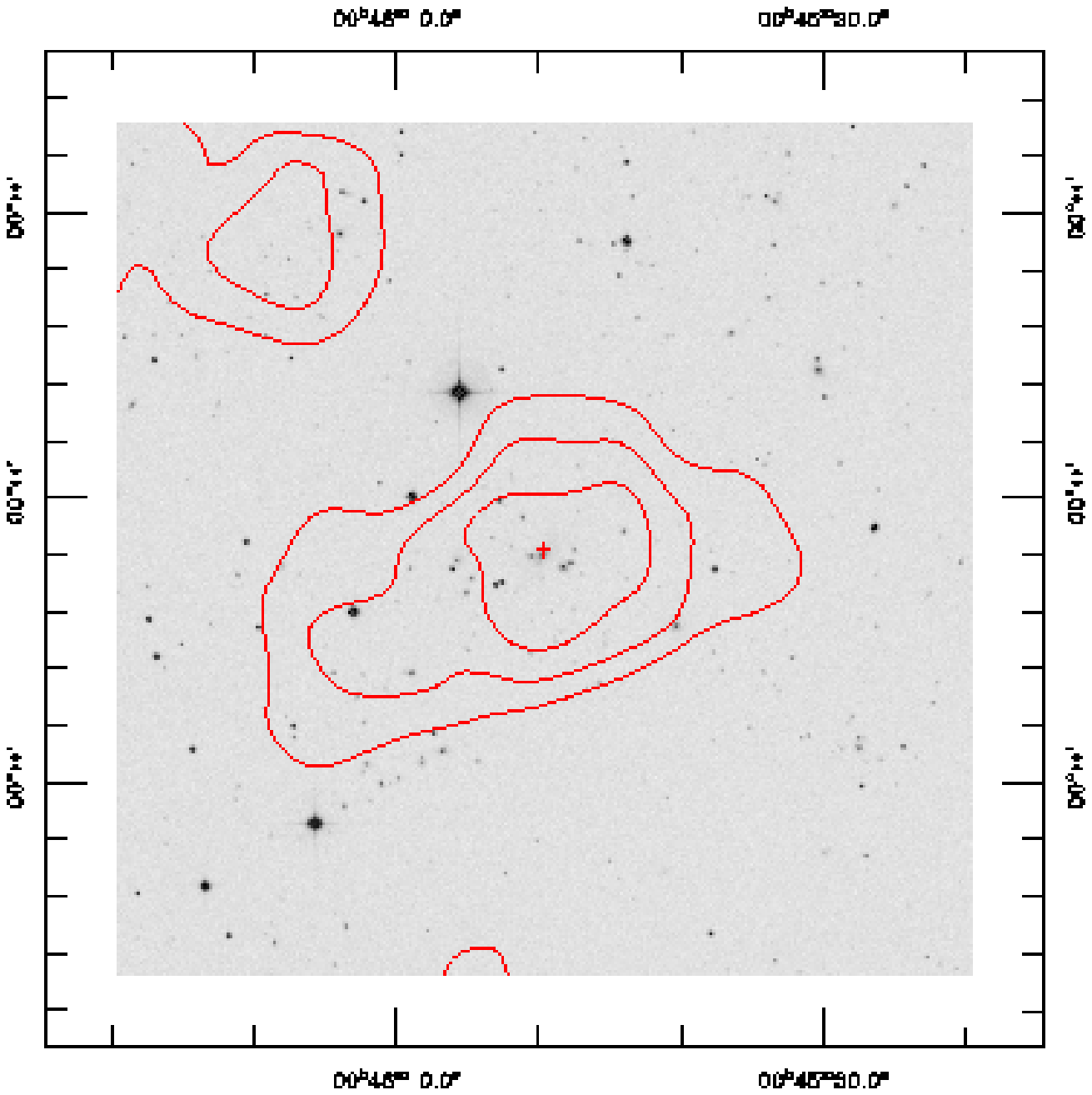,height=9.0cm,width=9.0cm}}
\vspace{-9.0cm}
\centerline{\hspace{9.0cm}
\psfig{figure=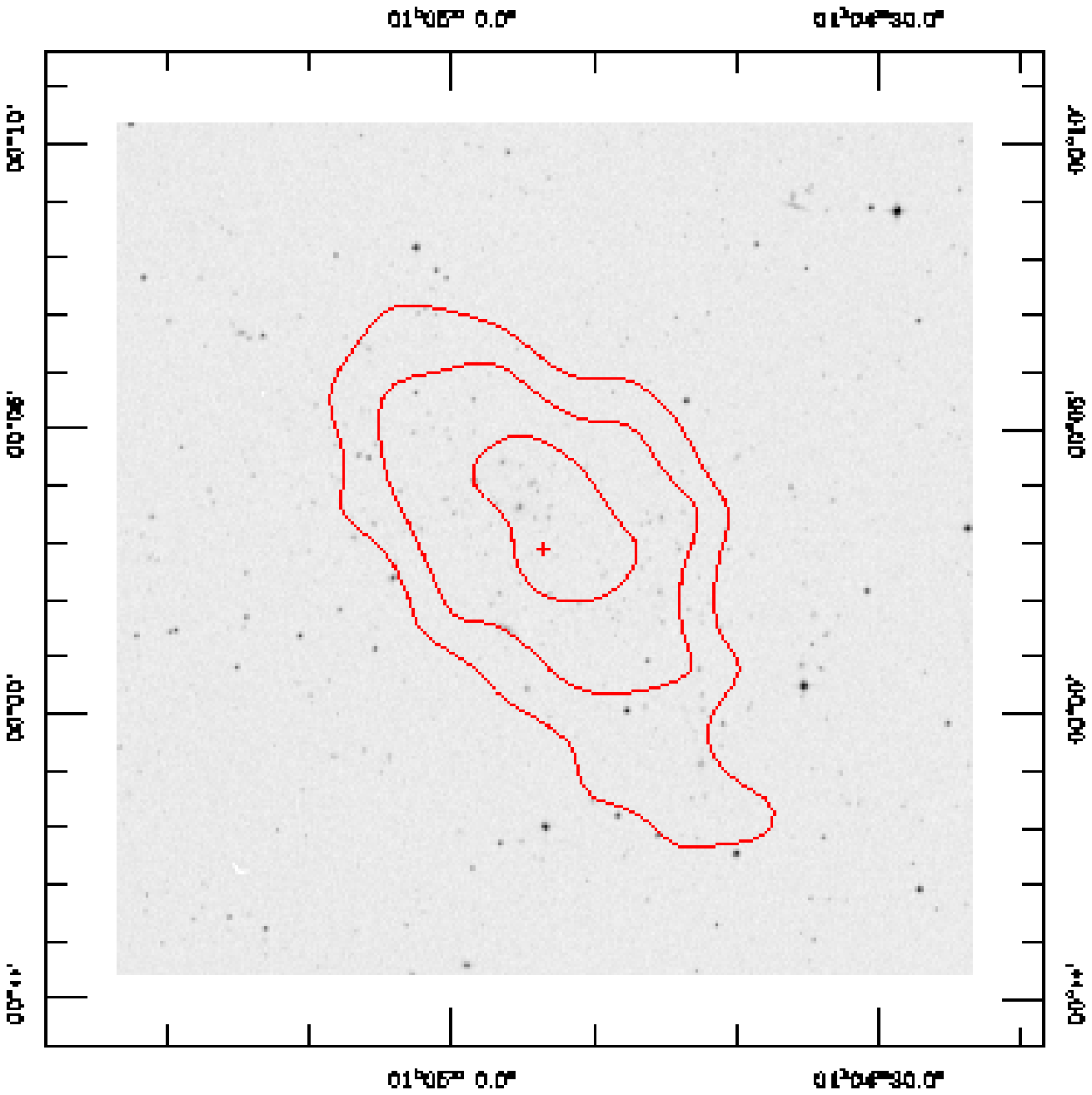,height=9.0cm,width=9.0cm}}
\vspace{1.4cm}
\caption{\small {\bf RS17} at $z=0.1092$ with the X-ray flux
$S=1.4\times 10^{-12}\,{\rm erg}\,{\rm s}^{-1}\,{\rm cm}^{-2}$ in the
ROSAT energy range 0.1-2.4\,keV. {\bf RS28} at $z=0.2745$ with
$S=1.3\times 10^{-12}\,{\rm erg}\,{\rm s}^{-1}\,{\rm cm}^{-2}$.}
\label{FIG_R1}
\end{figure*}

\begin{figure*}
\vspace{0.0cm}
\centerline{\hspace{-9.0cm}
\psfig{figure=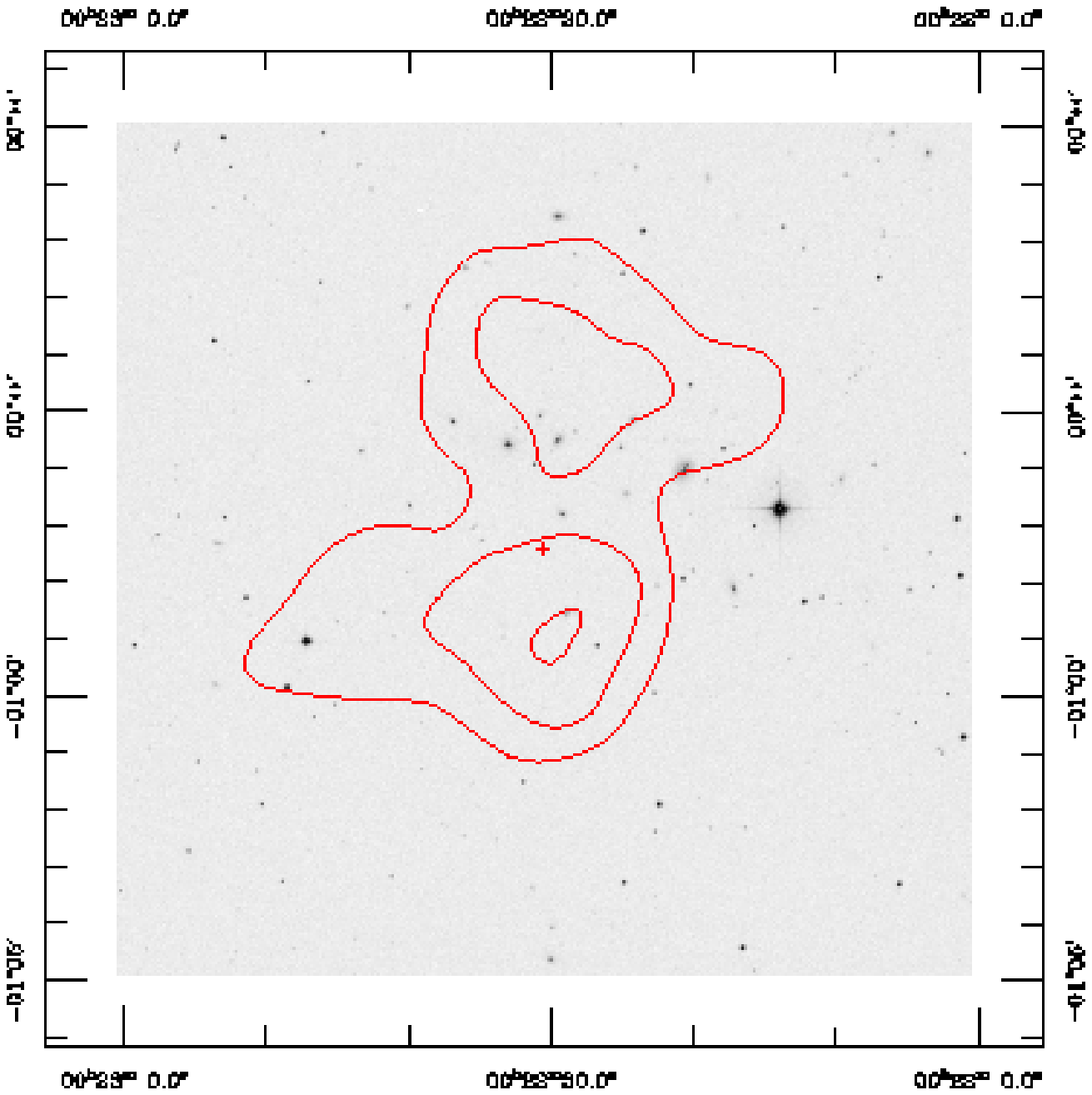,height=9.0cm,width=9.0cm}}
\vspace{-9.0cm}
\centerline{\hspace{9.0cm}
\psfig{figure=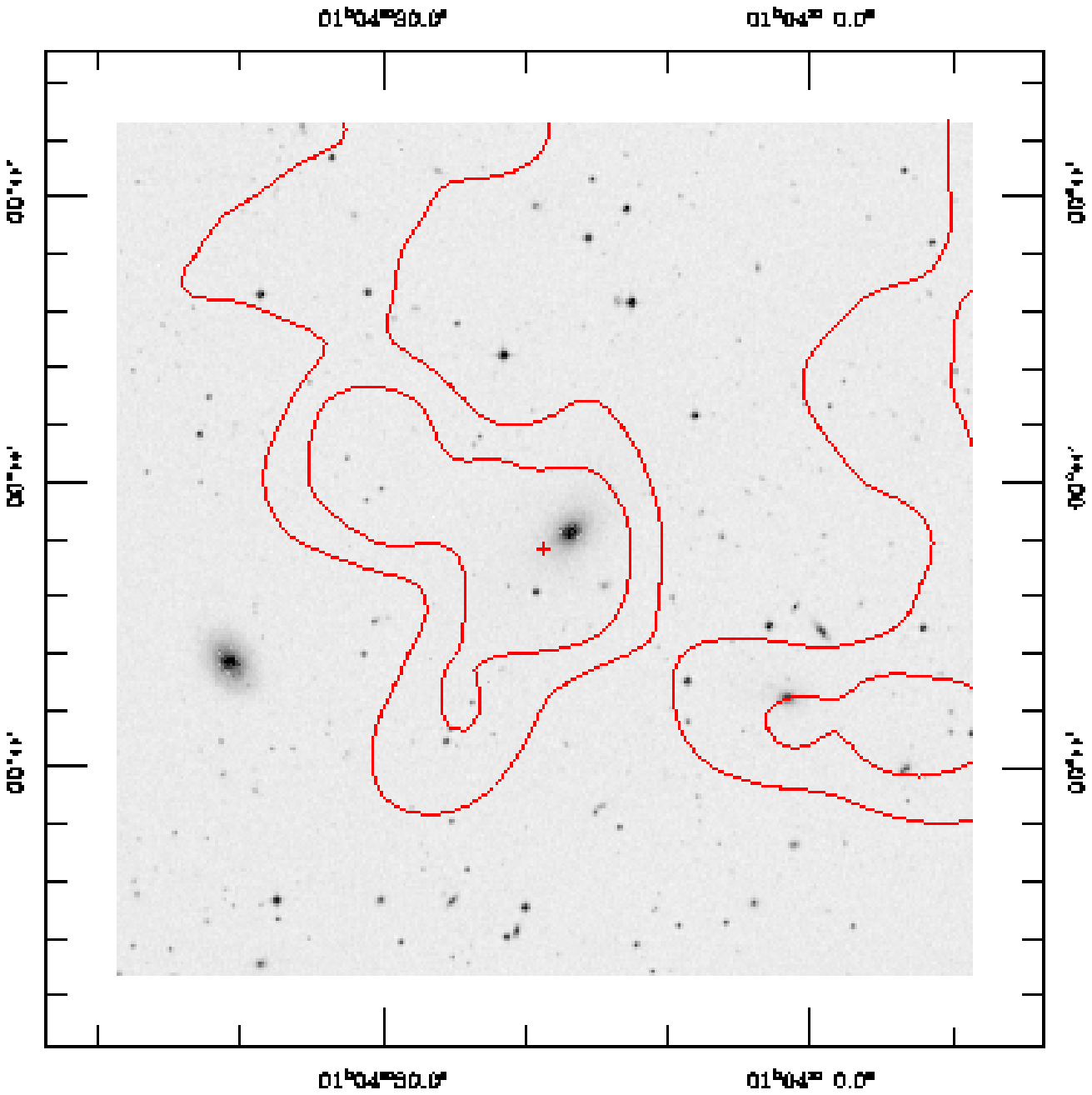,height=9.0cm,width=9.0cm}}
\vspace{+2.0cm}
\centerline{\hspace{-9.0cm}
\psfig{figure=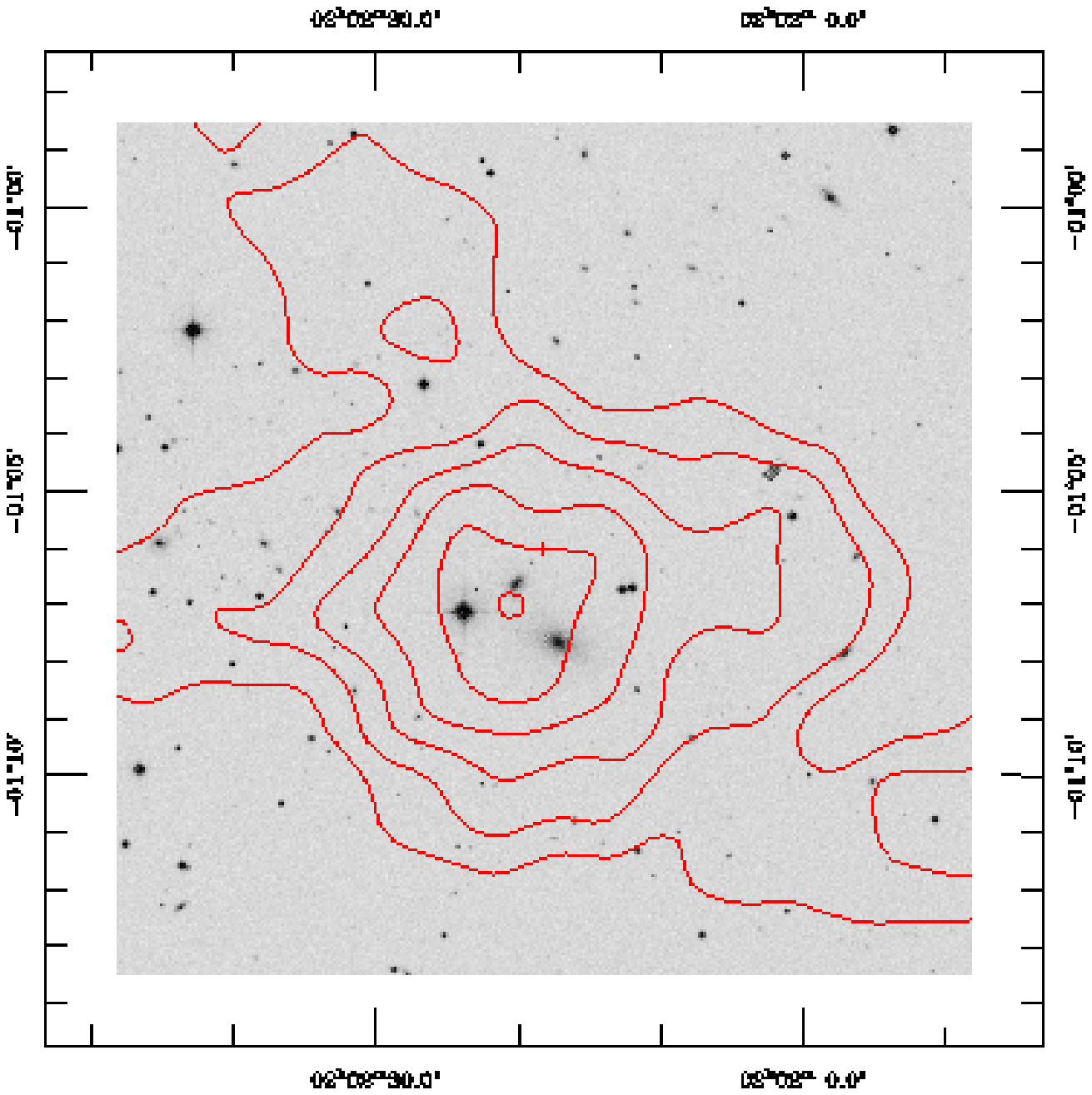,height=9.0cm,width=9.0cm}}
\vspace{-9.0cm}
\centerline{\hspace{9.0cm}
\psfig{figure=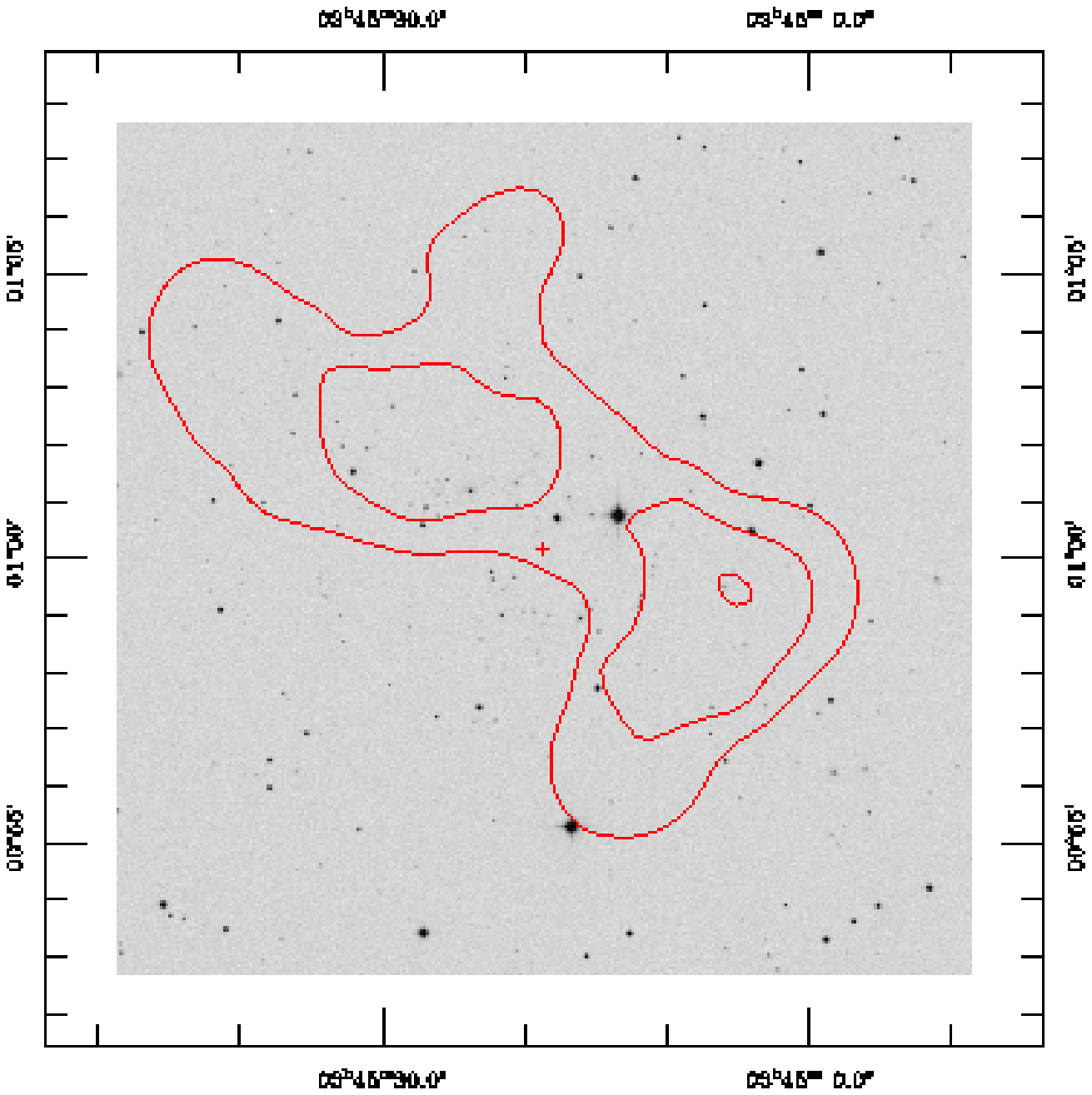,height=9.0cm,width=9.0cm}}
\vspace{1.4cm}
\caption{\small {\bf RS7} at $z=0.0581$ with the X-ray flux
$S=5.4\times 10^{-13}\,{\rm erg}\,{\rm s}^{-1}\,{\rm cm}^{-2}$ in the
ROSAT energy range 0.1-2.4\,keV. The cluster is not in the Bahcall et
al (2003) sample. {\bf RS26} at $z=0.0663$ with $S=5.1\times
10^{-13}\,{\rm erg}\,{\rm s}^{-1}\,{\rm cm}^{-2}$ is not in the
Bahcall et al. (2003) and Goto et al. (2002a) catalogues. {\bf RS53}
at $z=0.0425$ with $S=3.7\times 10^{-12}\,{\rm erg}\,{\rm
s}^{-1}\,{\rm cm}^{-2}$ is not in the Bahcall et al.  and Goto et
al. catalogues.  {\bf RS75} at $z=0.1822$ with $S=1.3\times
10^{-12}\,{\rm erg}\,{\rm s}^{-1}\,{\rm cm}^{-2}$ is not in the
Bahcall et al.  and Goto et al. catalogues. The cluster is also not in
the RBF sample extracted from RASS-2 (Voges et al. 1999).}
\label{FIG_R2}
\end{figure*}

In order to get more information about the reliability of the matched
filter countrates, we also compare with the countrates obtained from
ROSAT PSPC pointed observations as published in the 2nd ROSAT Source
Catalog of Pointed Observations with the Position Sensitive
Proportional Counter (2RXP, Fig.\,\ref{FIG_CTR}), independent whether
the target is an X-ray cluster or another X-ray source. The comparison
is not unrealistic because the majority (80\%, see below) of the
detected faint X-ray clusters appear point-like in RASS-3a. Averaged
systematic and random errors of the countrates obtained with RASS-3
and ROSAT PSPC pointed observations are shown in Table\,1.  We define
a characteristic lower $\Lambda$ limit by the value where the error of
the countrate has the same size as the countrate itself. In this case,
the comparison suggests a lower limit of about $\Lambda=0.04\,{\rm
s}^{-1}$.

Similarily, the systematic errors of the $\Lambda$ countrates as
estimated from the comparison with pointed observations range from an
underestimation of the countrates of $\sigma_\Lambda/\Lambda=3.9$\,\%
for $\Lambda>0.1\,{\rm s}^{-1}$ to an underestimation of 9.3\,\% at
$\Lambda>0.04\,{\rm s}^{-1}$. Below this countrate limit the
comparison suggests a possible overestimation of the countrates. One
could attribute this error to non-negligible contributions by
neighbouring X-ray sources. However, the systematics are of the order
of only 10\% compared to the random errors at these very low
countrates so that one should not overestimate the significance of
this effect.

After the countrate/flux conversion of the cluster candidates (see
below) we obtain the cumulative flux-number count histogram shown as
the continuous line in Fig.\,\ref{FIG_LOGN}. Only the raw counts
uncorrected for the effective survey area are shown so that the output
of the cluster search is seen more directly. The comparison shows that
our survey fills the gap between traditional large-area surveys like
REFLEX and faint serendipitous surveys like the RDCS (Rosati et
al. 1998) which have comparatively large statistical errors in this
range. A lower limit of the flux limit can be obtained from the
location where the histogram significantly flattens towards faint
fluxes (almost independent of the threshold likelihood of the SDSS
data). We estimated a nominal flux limit of $3-5\times 10^{-13}\,{\rm
erg}\,{\rm s}^{-1}\,{\rm cm}^2$ (0.1-2.4\,keV). Below this flux limit
a strong incompleteness is expected caused by, e.g., surface
brightness. A more quantitative analysis needs a larger test area as
provided by the SDSS Data Release 2 (coming soon) and a more detailed
work on simulated data (P. Schuecker et al., in preparation).

The comparison of the $z$ estimates obtained with the optical data
with spectroscopic redshifts shows no significant systematic
deviations and underlines the quality of our maximum likelihood
cluster redshift estimates (see Fig.\,\ref{FIG_ZZ} and Table\,2). Note
that X-ray clusters up to redshifts of at least $z=0.5$ can be
detected with our method with average redshift errors of about
$\sigma_z=0.03$.

Tables\,2 and 3 give a summary of the X-ray clusters selected from the
combination of RASS-3 and SDSS data. In Table\,2, column 1 contains a
running number of the X-ray clusters. Columns 2 and 3 give the cluster
sky coordinates (in degrees, Equinox 2000.0) computed by weighting
with the combined maximum likelihood maps of the X-ray and optical
data. Column 4 gives the column density of neutral galactic hydrogen
in units of $10^{20}\,{\rm cm}^{-2}$ as obtained from the 21\,cm
observation of Dickey \& Lockman (1990) and Stark et
al. (1992). Column 5 gives the unabsorbed X-ray flux in units of
$10^{-13}\,{\rm erg}\,{\rm s}^{-1}\,{\rm cm}^{-2}$ in the ROSAT energy
band 0.1-2.4\,keV determined for an assumed temperature of 5\,keV and
solar metallicity at redshift zero. For the countrate/flux conversion,
Tab.\,2 in B\"ohringer et al. (2003) was used. Columns 6 and 7 give
the maximum likelihood values obtained from RASS-3 and SDSS at the
final (R.A.,DEC.) positions, columns 8 and 9 the estimated cluster
redshifts obtained from the X-ray and optical data, respectively. The
cluster redshifts obtained from RASS-3 with values smaller than $z=1$
indicate the detection of an X-ray extent. A more precise measure of
source extent is given in Tab.\,3 (see below). Column 10 gives the
spectroscopic cluster redshift as found in the NASA/IPAC Extragalactic
Database (NED) and column 11 the number of cluster galaxies used to
determine the spectroscopic redshift. The cluster redshifts are
obtained from either a published cluster redshift or by the median of
several galaxies with measured redshifts. Galaxies are rejected when
they fall outside a 3000\,km/s velocity interval around the cluster
redshift. In some cases, the remark column in Table\,3 includes
additional information on $z$ in form of the cluster redshift $z_{\rm
G}$ from Goto et al. (2002a).  Column 12 indicates whether the X-ray
cluster is {\it not} included in several selected X-ray and cluster
source lists. The notation is as follows:

\begin{figure}
\vspace{-1.0cm}
\centerline{\hspace{0.0cm}
\psfig{figure=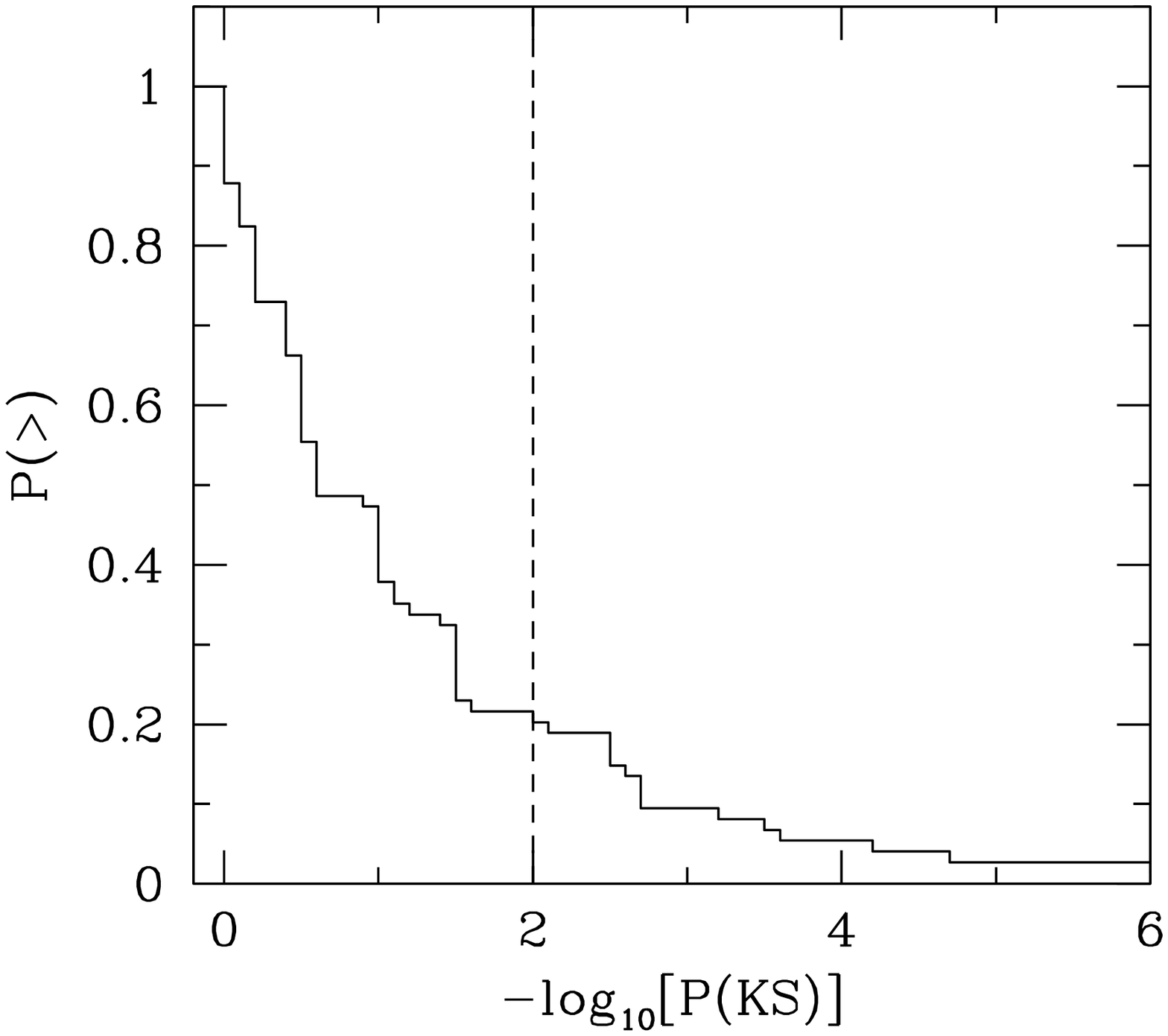,height=8.0cm,width=8.0cm}}
\vspace{-0.5cm}
\centerline{\hspace{0.0cm}
\psfig{figure=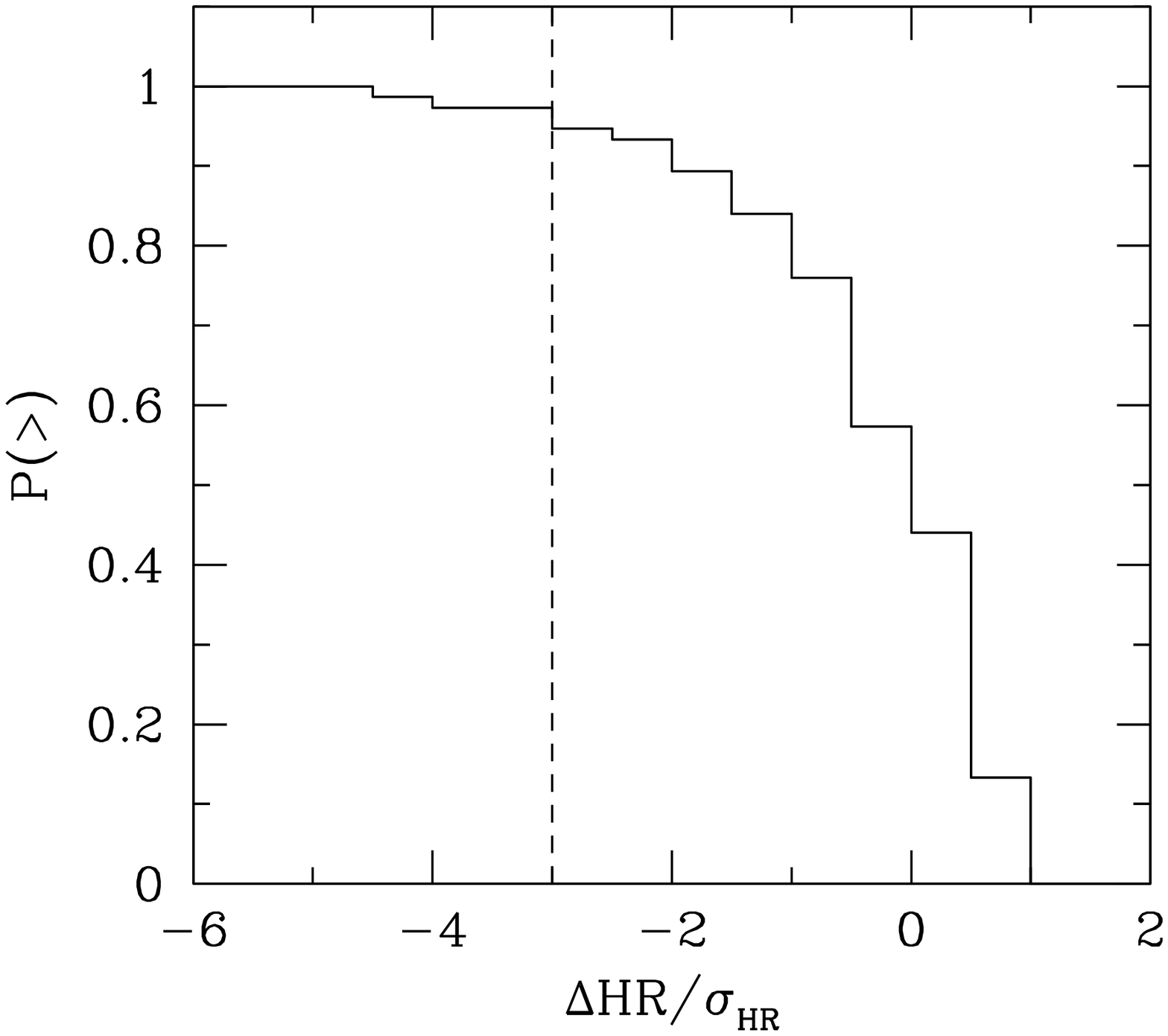,height=8.0cm,width=8.0cm}}
\vspace{-0.1cm}
\caption{\small {\bf Upper panel}: Cumulative distribution of the
negative logarithmic Kolmogorov-Smirnov probability for a point
source. The vertical line marks the threshold below which clusters are
expected to be point-like in RASS-3. {\bf Lower panel}: Cumulative
distribution of the normalized differences between the observed
hardness ratios to the theoretically expected hardness ratios for a
5\,keV cluster, normalized to the error of the hardness ratio
measurement. The vertical line marks the threshold below cluster
fluxes are expected to be significantly contaminated by AGN.  All
quantities are determined with the GCA method.}
\label{FIG_HREX}
\end{figure}

V$_{2}$: The ROSAT Bright and Faint Sources Catalogues based on RASS-2
(Voges et al. 1999); V$_3$: a preliminary (unscreened) MPE internal
source list based on RASS-3. For the detection of counterparts we
assume a search radius of 10 arcmin.

G: The SDSS cluster catalogue published in Goto et al. (2002a). For
the detection of counterparts we assume a search radius of 4 arcmin.

B: The SDSS cluster catalogue published in Bahcall et al. (2003). Note
that this list contains a subsample of a merged list of SDSS clusters
obtained with a Hybrid Matched Filter (Kim et al. 2002) and with a
color-magnitude red-sequence maxBCG technique (Annis et al. 2003). The
Bahcall et al. list has a threshold richness cut below the Abell
richness class 0 and accepts only clusters with redshifts
$0.05<z<0.3$. For the detection of counterparts we assume a search
radius of 4 arcmin.

Table\,3 gives more information about the selected X-ray
clusters. Columns 1-3 are the same as in Table\,2 and contain the
cluster number and the sky positions. In column 4 the source counts in
the ROSAT energy range 0.5-2.0\,keV at the final sky position are
given. Columns 5 and 6 contain the countrates determined with the
matched filter and with the GCA method, respectively. The formal
$1\sigma$ errors include the contribution from source count and
background. In column 8 the difference between the observed hardness
ratio and the theoretically expected hardness ratio as calculated for
a 5\,keV cluster and normalized to the error of the hardness ratio
measurement is given. In column 9 the negative logarithmic
Kolmogorov-Smirnow probabilities for a point source are given. The
hardness ratios and the extent probabilities are calculated with
GCA. The optical richness can be found in column 9. Special remarks
concerning the X-ray clusters are summarized in column 10.

The X-ray/optical overlays in Figs.\,\ref{FIG_R1}-\ref{FIG_R2} show
typical clusters of the present sample. The X-ray data are presented
in form of contours of constant signal-to-noise above background in
units of $1\sigma,2\sigma,\ldots$. The optical data are from digitized
Schmidt plates retrived from the 2nd version of the Digital Sky Survey
(DSS2) archive. The clusters RS26, RS53 and RS75 in Fig.\,\ref{FIG_R2}
are {\it not} in the SDSS cluster catalogues of Gogo et al. (2002a)
and Bahcall et al. (2003).

All clusters of the present sample show a concentration of optical
galaxies towards the X-ray center (in most cases clearly visible even
in the less deep DSS2 images). The selection criteria described in
Sect.\,\ref{RS} thus appear quite useful: all X-ray cluster candidates
passed the visual screening process. We have only excluded 7 targets
which were obvious fragments of already detected objects. They could
have been excluded already during the source detection in the maximum
likelihood maps by introducing a minimum distance between two detected
objects. However, from our past experience with NORAS and REFLEX we
prefer to reject objects only after a careful visual screening at the
end of the complete data processing when all data are in hand.

\section{Discussion and conclusions}\label{DISCUSS}

\begin{figure*}
\vspace{0.0cm}
\centerline{\hspace{-7.5cm}
\psfig{figure=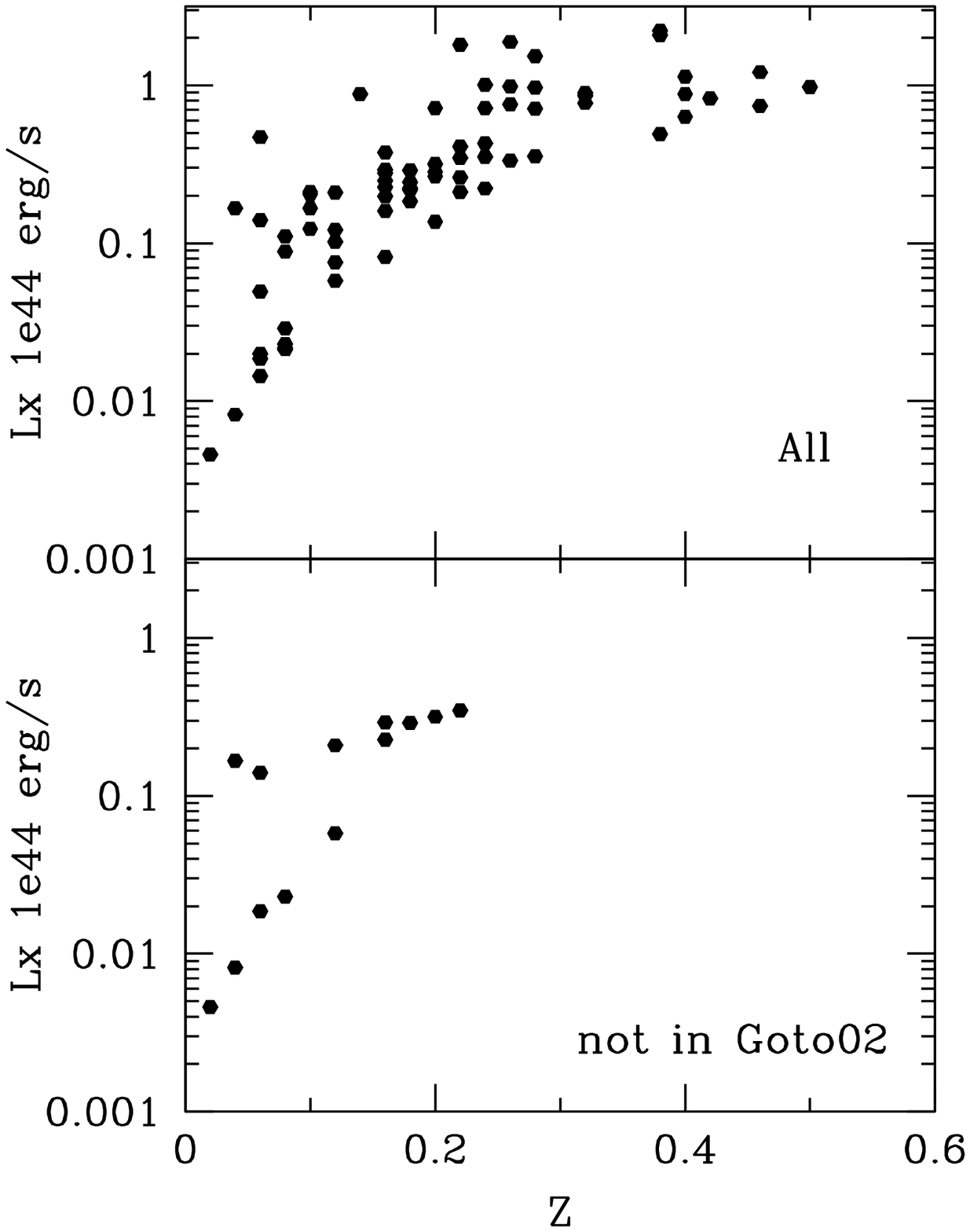,height=8.0cm,width=8.0cm}}
\vspace{-8.0cm}
\centerline{\hspace{10.5cm}
\psfig{figure=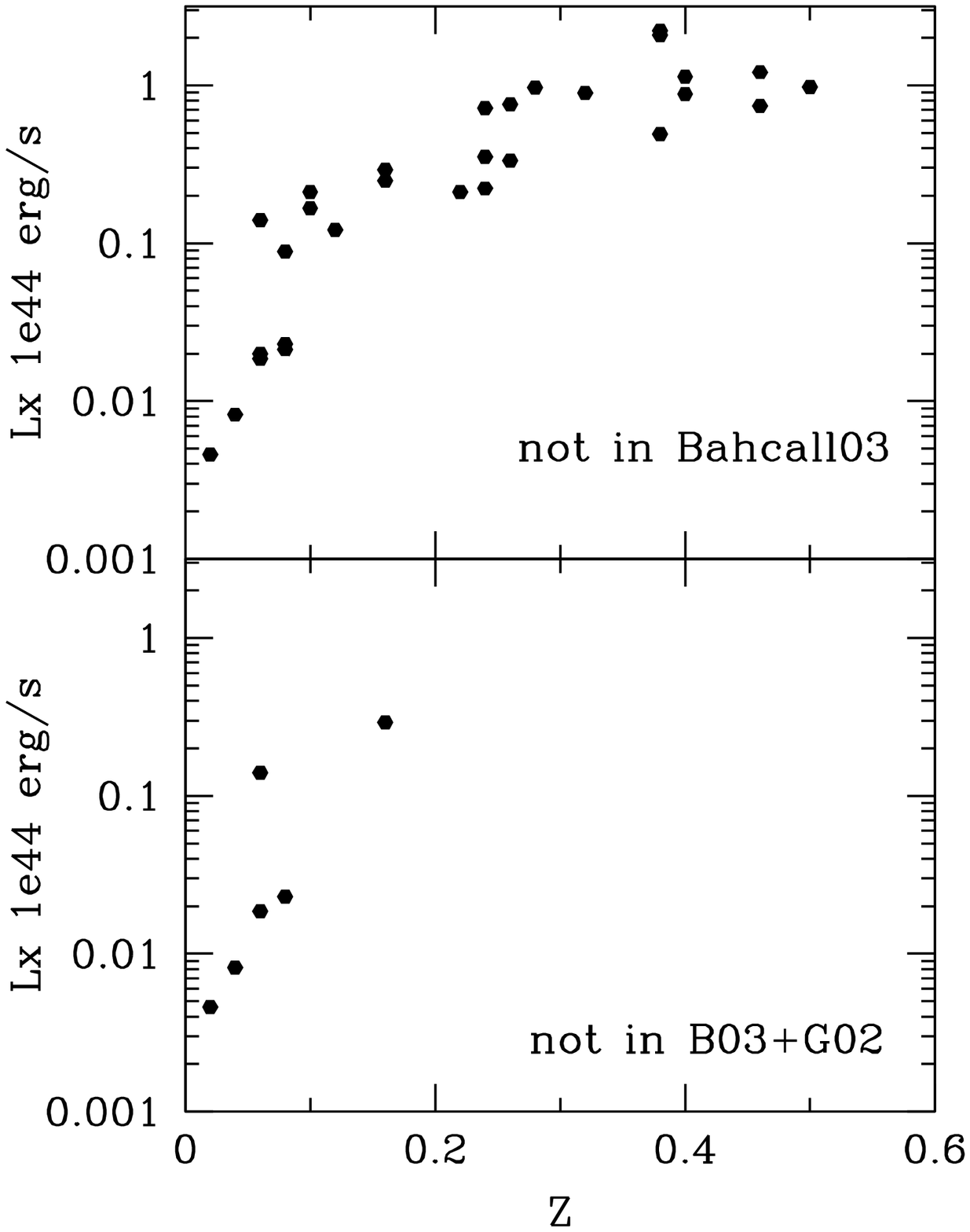,height=8.0cm,width=8.0cm}}
\vspace{0.0cm}
\caption{\small X-ray luminosity vs. redshift for the total cluster
sample (upper left), for clusters not in the Goto et al. (2002a)
sample (lower left), for clusters not in the Bahcall et al. (2003)
sample (upper right), and for the clusters neither in Goto et al. and
Bahcall et al. (lower right). The redshifts used are estimated from
the SDSS data with the matched filter described in Sect.\,\ref{SDSS}.}
\label{FIG_LZ}
\end{figure*}

The present paper describes a method to search for clusters of
galaxies by combining data obtained in different energy ranges. The
main goal is to reach lower flux limits and to improve the reliability
of a first identification of the sources. 

As an example, we combined X-ray data from RASS-3 and optical data
from SDSS. The SDSS galaxy sample reaches much higher redshifts than
RASS and can thus be used to guide a cluster detection in RASS down to
lower X-ray flux limits. We could show that a detection of X-ray
clusters in RASS-3 is possible down to a nominal X-ray flux limit of
$3-5\times 10^{-13}\,{\rm erg}\,{\rm s}^{-1}\,{\rm cm}^{-2}$ in the
ROSAT energy band 0.1-2.4\,keV. This limit is about 5--10 times deeper
than REFLEX and gives a first qualitative impression how much deeper
one can get with the new technique.

In the following we discuss the results which are already indicated by
the analysis of the test area. Table\,2 shows that only 40 out of our
75 X-ray clusters (53\%) are listed as X-ray sources in the ROSAT
Bright and Faint Source (RBF) catalogues. The overlap can be increased
possibly to 73\% when a preliminary source list obtained with RASS-3
is used. It is thus not only the X-ray flux limit which can be
decreased but also the X-ray completeness which can be increased when
RASS and SDSS data are combined.

At the nominal X-ray flux limit only 20\% of the 75 clusters show a
significant X-ray extent as measured with the GCA method
(Fig.\,\ref{FIG_HREX}, Tab.\,3). This small fraction is mainly caused
by the comparatively small number of source counts (half of the sample
has source counts between 5 and 10). Usually, more than 30 source
counts are necessary for the GCA method for a significant detection of
a source extent -- the precise numbers do, however, also depend on the
intrinsic extent of the source.

One could worry about a significant contamination by AGN. However,
only 12\% of the clusters have a known nearby AGN (Tab.\,3) and only
5\% of the X-ray clusters have an X-ray hardness ratio measured with
GCA which is consistent with an AGN (Fig.\,\ref{FIG_HREX}). The
fraction of contaminated clusters should thus be not very large,
possibly around 10\%.

Before we discuss the direct comparisons with published SDSS-based
cluster samples we want to note that it would not make much sense to
test the number of SDSS clusters which are not present in our sample
because SDSS-based cluster surveys are expected to reach much higher
redshifts and less rich systems compared to RASS-based cluster
surveys. We are thus mainly looking for X-clusters which are not found
in SDSS samples.

The comparison of our X-ray cluster sample with public SDSS cluster
samples yields the following results: 13 of our 75 X-ray clusters
(17\%) are not in the catalogue of Goto et al. (2002a) and 23 clusters
(31\%) are not in the catalogue of Bahcall et al. (2003). At the
comparatively bright X-ray fluxes $S\ge 5.0\cdot 10^{-13}\,{\rm
erg}\,{\rm s}^{-1}\,{\rm cm}^{-2}$ where spurious detections are
unlikely, still 11 clusters (15\% of our sample) are not in the Goto
et al. sample and 19 clusters (25\% of our sample) are not in the
Bahcall et al. sample.

However, when we combine the catalogues of Goto et al. and Bahcall et
al. then the incompleteness of the optical sample decreases to a level
of 8\%. The corresponding 6 X-ray clusters which are not in the two
SDSS-based cluster catalogues are marked as BG in the last column of
Table\,2 and have the spectroscopic redshifts $z=0.0637$, 0.0650,
0.0663, 0.0425, 0.0206 and 0.1822. Most of these missing clusters
(with one exception) are nearby and less rich systems. This is also
seen in the X-ray luminosity/redshift plots shown in
Fig.\,\ref{FIG_LZ}. In our X-ray cluster sample we have 12 clusters
with redshifts $z<0.07$ and only 7 (58\%) are in the combined optical
catalogues of SDSS. One should note that the 6 X-ray clusters missing
in the optical catalogues on our test area of 140 square degrees could
lead to an extrapolated number of 300 clusters over a total SDSS
survey area of 7000 square degrees. We are thus discussing here a
significant fraction of nearby systems.

\begin{figure}
\vspace{-2.2cm}
\centerline{\hspace{1.0cm}
\psfig{figure=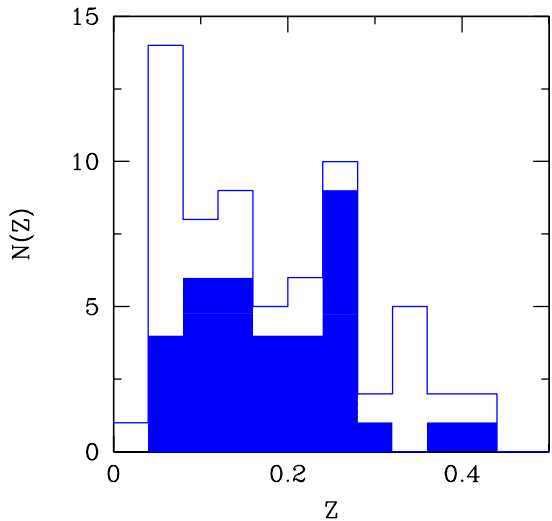,height=10.0cm,width=10.0cm}}
\vspace{-1.1cm}
\caption{\small Redshift histogram of the clusters with spectroscopic
redshifts. Filled steps show the redshift distribution of the clusters
which are in common with Goto et al. (2002a) and Bahcall et
al. (2003).}
\label{FIG_NZ}
\end{figure}

The difference between X-ray and optical cluster selection becomes
apparent in Fig.\,\ref{FIG_NZ}, where the redshift distribution of the
total sample (unfilled histogram) and the distribution of clusters
which are members of both the Goto et al. and the Bahcall et
al. samples are compared. Ignoring the redshifts $z>0.3$ where the
sample of Bahcall et al. is incomplete (by construction), the redshift
distributions are significantly different as seen by the strong
deficiency of nearby clusters in the combined Goto plus Bahcall
catalogue. Whereas the cluster counts of the combined optical
catalogue increase with redshift -- which is typical for
volume-limited samples -- the X-ray sample shows a steady decrease
with redshift -- which is typical for flux-limited samples. Numerical
simulations and comparisons with deep XMM and Chandra observations
will give further information about the final incompleteness level
which can be reached with our method.

We want to close the discussion of the sample with a few general
remarks on the quite deep X-ray flux limit achieved from the
combination of RASS and SDSS data. The resulting `superfaint' flux
limit -- though optimal for the detection of high-redshift clusters
and for compiling huge sample sizes -- yields X-ray cluster candidates
which consists in many cases of only 5--10 RASS X-ray source
counts. The corresponding formal flux errors are of the order of 30
and 40\%. A careful evaluation of the resulting accuracies achievable
in cosmological tests with such samples is thus necessary.

In addition, more work is needed to better calibrate and test our
method: (1) The analysis of a larger test area (Data Release 2) is
expected to give significantly smaller statistical errors of the
cluster number counts which is necessary for a better calibration of
the likelihood threshold of the optical data. (2) Tests are under way
which quantify the effect an optical sample like SDSS has on the
selection of X-ray clusters and vice versa. In Popesso et al. (2004)
the correlations between basic optical and X-ray cluster properties
are studied and more tests will follow. One of their conclusions is
that the $z'$ SDSS photometric band and not the $r'$ band used here
gives the strongest correlations between optical and X-ray data, and
should thus be the preferred band when we want to work with the
smallest bias caused by including optical data in the detection
process. (3) The numerical simulations in Schuecker \& B\"ohringer
(1998) have to be performed under the specific conditions of the
present survey in order to test selection effects caused especially by
the small number of X-ray photons and surface brightness. (4) A direct
comparison with ROSAT, XMM-Newton, and Chandra pointed observations
will give model-independent estimates of the flux and completeness
limits.

\subsection{Role of Virtual Observatories}\label{ROLE}

The present investigation gives a good illustration of the benefits
achievable from the combination of high-quality optical and X-ray
data. The combination of the two large databases allows us to fill the
gap between nearby wide-angle and deep serendipitous X-ray cluster
surveys with already existing data. The combination increases the
completeness of the samples and reduces the time-consuming
spectroscopic follow-up observations. This method is also expected to
be helpful when samples of the order of several $10^4$ X-ray cluster
candidates expected from future wide-angle X-ray surveys like DUO or
ROSITA must be identified.

However, the quality and completeness of the catalogues depends also
on the model used to detect and characterize the galaxy clusters. In
the present investigation only one set of values of model parameters
(cluster core radius, cutoff radius, Schechter characteristic
magnitude and faint end slope) was used to develope and illustrate the
method. More realistic applications should test different sets of
model parameters and use a finer sampling grid for cluster
detection. In addition, cluster radii are known to increase with mass
so that some kind of iterative procedure is necessary to find the
optimal model for each cluster.

Further applications should also include the detection of clusters in
so-called Sunyaev-Zel'dovich maps obtained in the centimeter or
millimeter range. Such kinds of maps will be provided by, e.g., the
PLANCK mission, and will surely increase the reliability of the
resulting catalogues.

We are thus left with the conclusion that the quality, the
completeness limit, the detection limit, and the telescope time for
source identification can be optimized when enough computer power to
sample the high-dimensional parameter spaces is available. Moreover, a
fast and easy handling of data is needed to work with very large
databases probably located at different physical nodes. {\it Virtual
observatories} like the GAVO provide the infrastructure for such kind
of investigation and are surely necessary to make full and efficient
use of already existing and future observed data.

GAVO provides an all-sky X-ray event file including many
characteristics of the invididual photons. The photons are retrieved
from the archive with a general cone-search algorithm. We are
currently in the process to implement the cluster search described
here within the GAVO grid. This final step is crucial for the analyses
of larger survey areas with more cluster models and a finer sampling
grid of the detection cells. This will allow us to analyse in a much
more efficient manner the SDSS Data Release 2 sample which will become
available soon and RASS-3.

The method described in the present investigation needs about 1 hour
computing time per square degree on a standard workstation with one
processor. Running the method in a controlled manner over a grid of
many computers is thus essential for a proper analysis within
reasonable times. Our plans include the provision of the cluster
(source) detection algorithm as a service of GAVO so that the
procedures can be used in a user-friendly manner by a larger community
of interested researchers.

\begin{acknowledgements}
We would like to thank the ROSAT team at MPE for the support for the
data reduction of the ROSAT All-Sky Survey and the ROSAT XRT archive
of pointed observations, and the REFLEX team for their help in the
preparation of the REFLEX cluster sample. We further thank the GAVO
team for technical support and for the possibility to run the jobs on
the GAVO computer network. We also thank the anonymous referee for
useful comments. The Sloan Digital Sky Survey is a joint project of
the University of Chicago, Fermilab, the Institute of Advanced Study,
the Japan Participation Group, Johns Hopkins University, the Max
Planck Insitut f\"ur Astronomie, the Max Planck Institut f\"ur
Astrophysik, New Mexico State University, Princeton University, the US
Naval Observatory, and the University of Washington. Apache Point
Observatory, site of the SDSS telescopes, is operated by the
Astrophysical Research Consortium. Funding for the project has been
provided by the Alfred P. Sloan Foundation, the SDSS member
institutions, the National Science Foundation, the US Department for
Energy, the Japanese Monbukagakusho, and the Max Planck
Gesellschaft. This research also made use of the NASA/IPAC
Extragalactic Database (NED), which is operated by the Jet Propulsion
Laboratory, California Instiute of Technology, under contract with
NASA. P.S. acknowledges support under the DLR grant
No.\,50\,OR\,9708\,35.
\end{acknowledgements}

\begin{table*}
{\bf Tab.\,2.} Catalogue of RASS-3/SDSS cluster candidates. Col.\,1:
Cluster number. Col.\,2-3: cluster coordinates in degrees. Col.\,4:
column density of neutral galactic hydrogen in units of $10^{20}\,{\rm
cm}^{-2}$. Col.\,5: unabsorbed X-ray flux in units of $10^{-13}\,{\rm
erg}\,{\rm s}^{-1}\,{\rm cm}^{-2}$ in the ROSAT energy band
0.1-2.4\,keV. Col.\,6-7: detection likelihoods on RASS-3 photon maps
and SDSS galaxy maps. Col.\,8-9: estimated cluster redshifts from
RASS-3 and SDSS data. Col.\,10: Spectroscopic cluster
redshift. Col.\,11: Number of cluster galaxies (after rejection of
outliners) used for the determination of the spectroscopic cluster
redshift. Col.\,12: Note whether the cluster is {\it not} found in the
RASS source catalogues (V), and not found in the SDSS cluster
catalogues compiled by Bahcall et al. (B, 2003) and Goto et al. (G,
2002a). For more details see main text.\\
\vspace{-0.1cm}
\begin{center}
\begin{tabular}{rrrrrrrrrrrr}
\hline
\hline
No.  &  R.A. &   DEC. & $n_{\rm H}$ & $S$ & $\ln(L/L_0)$ &  $\ln(L/L_0)$ & $z_{\rm est}$ & $z_{\rm est}$ &  $z_{\rm spec}$& $N_z$& Not in  \\
     & 2000.0& 2000.0 &             &     & RASS         &  SDSS         & RASS          & SDSS          &                &      & catalogue  \\
\hline
RS01 & 0.3581& -0.0312& 3.24&   7.39&  6.775      &   19.705      &    0.40       &    0.32       &    0.2578      &    3 & \\
RS02 & 4.3972& -0.8836& 2.95&   6.70&  7.645      &   38.440      &    1.00       &    0.24       &    0.2111      &      & \\
RS03 & 5.0566&  0.1015& 2.75&   7.21&  8.094      &   73.068      &    0.42       &    0.20       &    0.2100      &    5 & G\\
RS04 & 5.1680& -0.9858& 2.85&   5.01&  5.202      &   15.987      &    1.00       &    0.06       &    0.0637      &    4 & V$_2$BG\\
RS05 & 5.1943& -1.0844& 2.85&   3.44&  3.184      &   15.844      &    1.00       &    0.08       &    0.0650      &    6 & V$_2$BG\\
RS06 & 5.4073& -0.8459& 2.81&   4.34&  4.007      &   77.041      &    1.00       &    0.08       &    0.1042      &    5 & \\
RS07 & 5.6274& -0.9570& 2.81&   5.35&  5.408      &   29.057      &    1.00       &    0.06       &    0.0581      &   10 & V$_2$B\\
RS08 & 5.7515& -0.1543& 2.73&   6.17&  7.786      &  102.499      &    1.00       &    0.18       &    0.1582      &    9 & \\
RS09 & 5.7988&  0.1977& 2.67&   7.62&  3.360      &   18.930      &    0.10       &    0.22       &                &      & V$_2$\\
RS10 & 7.1309& -0.0857& 2.66&   3.85&  4.616      &   19.018      &    1.00       &    0.06       &                &      & V$_2$\\
RS11 & 7.3730& -0.2509& 2.66&  16.44& 10.839      &   60.477      &    0.06       &    0.08       &    0.0602      &   17 & \\
RS12 & 8.0414& -0.6574& 2.70&   3.10&  4.327      &   19.789      &    1.00       &    0.20       &                &      & \\
RS13 & 8.5759&  0.8777& 2.46&   6.39&  8.643      &   38.450      &    0.56       &    0.20       &    0.1910      &    3 & \\
RS14 & 9.3587&  0.1414& 2.26&   3.73&  3.384      &   21.011      &    1.00       &    0.12       &    0.0775      &    4 & G\\
RS15 &10.7036&  0.2149& 2.20&   7.87&  6.560      &   50.102      &    0.24       &    0.28       &    0.2688      &    2 & V$_{23}$\\
RS16 &10.9596&  0.1187& 2.18&   5.12&  4.611      &   27.796      &    1.00       &    0.18       &    0.2168      &    4 & V$_{23}$\\
RS17 &11.4570& -0.8486& 2.83&  13.66& 18.840      &   26.153      &    0.58       &    0.12       &    0.1092      &    8 & G\\
RS18 &11.5751&  0.0167& 2.23&   6.57&  7.329      &   51.584      &    1.00       &    0.12       &    0.1129      &   15 & \\
RS19 &11.6170& -0.2219& 2.34&   6.23&  6.697      &   24.532      &    0.78       &    0.18       &    0.1129      &    5 & V$_{23}$\\
RS20 &11.8959& -0.8947& 2.86&   7.15&  4.906      &   21.918      &    0.18       &    0.16       &    0.1178      &   10 & V$_{23}$\\
RS21 &13.9519&  0.6488& 2.84&  15.83& 11.479      &   37.177      &    0.08       &    0.10       &    0.0695      &   11 & B\\
RS22 &14.0756& -1.1735& 3.17& 127.50&215.523      &   45.265      &    0.08       &    0.06       &    0.0447      &   24 & \\
RS23 &15.6415& -0.8327& 3.68&   4.04&  3.809      &   26.316      &    0.92       &    0.22       &    0.2408      &    4 & B\\
RS24 &15.6763&  1.1263& 3.09&  15.81& 28.725      &   60.191      &    0.24       &    0.24       &    0.1437      &    6 & \\
RS25 &15.9883& -0.8067& 3.46&   3.31&  3.046      &   30.689      &    1.00       &    0.08       &    0.0662      &    9 & \\
RS26 &16.0781& -0.7695& 3.38&   5.08&  6.110      &   20.244      &    1.00       &    0.04       &    0.0663      &    7 & BG\\
RS27 &16.0791& -0.4288& 3.37&   8.30&  9.160      &   41.826      &    0.24       &    0.16       &                &      & G\\
RS28 &16.2231&  0.0484& 3.42&  13.15& 20.301      &   88.368      &    0.24       &    0.26       &    0.2745      &    4 & \\
RS29 &16.3765& -0.1699& 3.32&   7.26&  4.274      &   17.245      &    0.10       &    0.20       &    0.2480      &    2 & V$_{23}$\\
RS30 &16.5541&  0.8009& 3.19&  25.08&  5.004      &   25.453      &    0.02       &    0.26       &    0.2620      &    3 & V$_{23}$\\
RS31 &16.7056&  1.0513& 3.15&  33.94&124.722      &   32.759      &    1.00       &    0.22       &    0.2545      &      & \\
RS32 &16.7805& -0.3280& 3.34&   6.67&  7.260      &   39.992      &    0.24       &    0.32       &    0.2802      &    2 & \\
RS33 &17.8626&  0.5741& 3.08&   3.38&  4.330      &   19.399      &    1.00       &    0.40       &    0.4066      &    2 & V$_{23}$\\
RS34 &18.7749&  0.3502& 3.38& 103.19&128.658      &   66.772      &    0.04       &    0.04       &    0.0449      &   27 & G\\
RS35 &20.5041&  0.3503& 3.30&  10.18& 19.504      &   47.098      &    1.00       &    0.16       &    0.1751      &    4 & \\
RS36 &21.8232&  0.3523& 3.12&  13.25& 27.011      &   51.312      &    0.52       &    0.38       &    0.3437      &    2 & B\\
RS37 &22.0578& -0.6777& 3.37&   4.07&  5.508      &   18.410      &    1.00       &    0.28       &    0.2566      &    1 & V$_{23}$\\
RS38 &22.6058&  0.4817& 2.98&   2.92&  3.280      &   19.801      &    1.00       &    0.38       &    0.3365      &    3 & V$_{23}$B\\
RS39 &22.9211&  0.5533& 2.93&  19.44& 24.895      &   28.408      &    0.12       &    0.10       &    0.0805      &   11 & \\
RS40 &23.7341& -0.9379& 3.07&   3.22&  3.001      &   18.683      &    0.64       &    0.08       &    0.0795      &    4 & B\\
RS41 &24.3671& -0.1861& 2.89&   4.77&  4.817      &   29.844      &    1.00       &    0.46       &                &      & V$_2$B\\
RS42 &26.6947& -0.6782& 2.92&  20.10& 59.417      &   20.039      &    1.00       &    0.10       &    0.0825      &   10 & B\\
RS43 &27.2777& -0.4271& 2.87&  10.01&  5.210      &   17.457      &    0.06       &    0.26       &    0.3336      &    1 & V$_{23}$B\\
RS44 &27.3363& -1.1850& 2.84&   3.22&  4.958      &   17.805      &    1.00       &    0.50       &                &      & B\\
RS45 &27.3458& -0.3773& 2.86&   7.64&  4.323      &   16.164      &    0.08       &    0.32       &    0.3336      &    1 & V$_{23}$B\\
RS46 &27.8459& -1.0133& 2.76&   2.95&  3.502      &   16.051      &    1.00       &    0.16       &    0.1209      &    3 & V$_2$\\
RS47 &28.1781&  1.0044& 2.80&  41.83&102.653      &  138.260      &    0.20       &    0.14       &    0.2300      &      & \\
RS48 &28.6185& -0.6531& 2.60&   4.63&  7.332      &   20.698      &    1.00       &    0.40       &    0.0463      &    6 & B\\
RS49 &29.0463&  1.0481& 2.87&  11.72& 13.089      &   86.197      &    0.16       &    0.10       &    0.0794      &   22 & V$_2$\\
\hline
\hline
\end{tabular}
\end{center}
\end{table*}

\begin{table*}
{\bf Tab.\,2.} Continued\\
\vspace{-0.1cm}
\begin{center}
\begin{tabular}{rrrrrrrrrrrr}
\hline
\hline
No.  &  R.A. &   DEC. & $n_{\rm H}$ & $S$   &$\ln(L/L_0)$ &  $\ln(L/L_0)$ & $z_{\rm est}$ & $z_{\rm est}$ &  $z_{\rm spec}$& $N_z$& Not in  \\
     & 2000.0& 2000.0 &             &       &RASS         &  SDSS         & RASS          & SDSS          &                &      & catalogue  \\
\hline
RS50 &29.1534&  0.8521&  2.79&  6.06& 10.140      &   47.039      &    1.00       &    0.20       &    0.2187      &    4 & \\
RS51 &29.3443& -0.1020&  2.63&  3.93&  4.010      &   33.930      &    1.00       &    0.42       &    0.3910      &    2 & V$_2$\\
RS52 &29.4179& -0.1211&  2.62&  6.44&  4.329      &   36.045      &    0.14       &    0.22       &    0.1352      &    5 & V$_2$G\\
RS53 &30.5762& -1.1003&  2.61& 37.46& 60.376      &   22.608      &    0.10       &    0.06       &    0.0425      &   15 & BG\\
RS54 &32.7195& -1.1530&  2.55&  6.87&  6.466      &   24.592      &    0.28       &    0.18       &    0.1731      &    5 & V$_{23}$\\
RS55 &33.5548& -0.1419&  2.76&  5.77&  5.717      &   16.778      &    1.00       &    0.16       &    0.1427      &    6 & \\
RS56 &34.6231& -0.4454&  2.96&  5.50&  3.012      &   54.774      &    1.00       &    0.24       &    0.2910      &    2 & V$_{23}$B\\
RS57 &37.0257& -1.1727&  2.68& 13.24& 10.988      &   16.801      &    0.20       &    0.08       &    0.0690      &   14 & V$_{23}$B\\
RS58 &38.9997& -1.0232&  2.85& 17.24&  4.707      &   16.673      &    0.12       &    0.28       &    0.2470      &    2 & V$_{23}$\\
RS59 &41.4703& -0.6960&  4.02& 14.06&  5.015      &   74.357      &    1.00       &    0.16       &    0.1815      &    6 & V$_2$\\
RS60 &42.0287& -0.6403&  4.43& 11.75&  4.911      &   20.581      &    1.00       &    0.02       &    0.0206      &      & V$_{23}$BG\\
RS61 &42.3519&  1.0279&  5.05& 13.02& 10.612      &   18.679      &    1.00       &    0.38       &                &      & V$_2$B\\
RS62 &46.0947&  1.0847&  7.60&  5.60&  5.436      &   46.489      &    1.00       &    0.12       &    0.1548      &      & V$_{23}$\\
RS63 &46.1463& -0.8985&  7.02& 10.01& 10.328      &   34.808      &    1.00       &    0.16       &    0.1586      &    3 & B\\
RS64 &46.5799& -0.2233&  4.99& 14.03&  4.861      &   15.094      &    0.06       &    0.06       &    0.1080      &   11 & V$_{23}$\\
RS65 &47.9288&  1.1583&  8.16&  3.97&  4.752      &   25.775      &    1.00       &    0.24       &    0.2130      &    1 & B\\
RS66 &50.1187&  0.5485&  7.21& 12.23& 10.466      &   23.010      &    0.20       &    0.28       &    0.3851      &    3 & B\\
RS67 &52.0082&  0.4007&  7.72&  4.99&  4.947      &   56.181      &    1.00       &    0.26       &    0.3202      &    3 & V$_{23}$B\\
RS68 &52.1520& -0.0535&  7.60& 12.60& 21.585      &   16.214      &    1.00       &    0.24       &                &      & B\\
RS69 &52.6521& -0.5483&  7.31&  8.90&  8.174      &   19.762      &    0.22       &    0.12       &    0.1487      &    4 & V$_2$B\\
RS70 &53.1336&  0.3992&  7.73&  3.30&  3.114      &   19.936      &    1.00       &    0.46       &    0.4256      &    3 & V$_2$B\\
RS71 &53.5782& -1.1857&  7.49& 18.42& 45.316      &   33.061      &    1.00       &    0.20       &    0.1387      &    8 & \\
RS72 &53.6314& -0.7342&  7.88&  9.35& 12.491      &   51.521      &    0.34       &    0.18       &                &      & G\\
RS73 &53.9795& -0.6320&  8.01&  5.58&  4.455      &   20.131      &    0.22       &    0.22       &    0.2741      &    2 & V$_{23}$\\
RS74 &55.6623& -0.2708&  7.79&  6.83&  5.461      &   42.824      &    0.28       &    0.40       &                &      & V$_2$B\\
RS75 &56.3280&  1.0026& 10.60& 12.86&  6.229      &   27.567      &    0.08       &    0.16       &    0.1822      &    4 & V$_2$BG\\
\hline
\hline
\end{tabular}
\end{center}
\end{table*}

\begin{table*}
{\bf Tab.\,3.} Catalogue of RASS-3/SDSS cluster candidates. Col.\,1:
cluster number. Cols.\,2-3: cluster coordinates in degrees. Col.\,4:
source counts. Col.\,5: countrates from matched filter and formal
$1\sigma$ errors. Col.\,6: countrate from Growth Curve Analysis GCA
and formal $1\sigma$ errors. Col.\,7: normalized hardness ratio (see
text). Col.\,8: negative logarithmic Kolmogorov-Smirnov probability
for a point source. Col.\,9: optical richness. Col.\,10: further
remarks (CONT: AGN within 1\,arcmin from cluster position in NED; 
CUT: cluster only partially in test area of matched filter).\\
\vspace{-0.1cm}
\begin{center}
\begin{tabular}{rrrrccrrrl}
\hline
\hline
No.  &  R.A. &   DEC. & $N_{\rm ph}$ & $\Lambda$        &  ctr(GCA)    & HR & EXT & Richness & Remarks \\
     & 2000.0& 2000.0 &              & $[{\rm s}^{-1}]$ & $[{\rm s}^{-1}]$  &GCA & GCA & SDSS     &         \\ 
\hline		       
RS01 & 0.3581& -0.0312&     12.5     & 0.036$\pm0.010$  & 0.021$\pm0.010$   &  0.59 & 1.4& 79.0   & \\    
RS02 & 4.3972& -0.8836&     11.3     & 0.033$\pm0.010$  & 0.023$\pm0.011$   &  0.75 & 0.4& 90.3   & ZwCl 0015.1-0108\\
RS03 & 5.0566&  0.1015&     13.9     & 0.036$\pm0.010$  & 0.026$\pm0.009$   & -1.46 & 0.2&128.0   & RXC00201.1\\
RS04 & 5.1680& -0.9858&      8.8     & 0.025$\pm0.008$  & 0.018$\pm0.008$   & -1.52 & 0.0& 41.4   & 2nd CLST $z=0.1071$\\   
RS05 & 5.1943& -1.0844&      6.0     & 0.017$\pm0.007$  & 0.015$\pm0.007$   &  0.57 & 2.1& 37.0   & \\
RS06 & 5.4073& -0.8459&      8.0     & 0.021$\pm0.007$  & 0.015$\pm0.007$   & -0.27 & 1.4&113.3   & A0023\\
RS07 & 5.6274& -0.9570&     10.2     & 0.026$\pm0.008$  & 0.026$\pm0.015$   &  0.80 & 0.0& 52.3   & $z_{\rm G}=0.12921$\\
RS08 & 5.7515& -0.1543&     12.1     & 0.030$\pm0.009$  & 0.022$\pm0.010$   &  0.02 & 0.1&151.3   & A0025\\ 
RS09 & 5.7988&  0.1977&     14.5     & 0.038$\pm0.010$  & 0.022$\pm0.010$   & -0.30 & 1.0& 62.1   & $z_{\rm G}=0.25402$\\
RS10 & 7.1309& -0.0857&      9.1     & 0.019$\pm0.006$  & 0.013$\pm0.006$   &  0.51 & 0.6& 46.8   & ZwCl 0025.9-0.021, CONT\\
RS11 & 7.3730& -0.2509&     39.7     & 0.081$\pm0.013$  & 0.082$\pm0.018$   & -1.11 & 1.5& 92.0   & ZwCl 0027.0-0036\\ 
RS12 & 8.0414& -0.6574&      9.7     & 0.015$\pm0.005$  & 0.009$\pm0.005$   & -0.54 & 0.2& 56.5   & $z_{\rm G}=0.24267$\\
RS13 & 8.5759&  0.8777&     13.6     & 0.032$\pm0.009$  & 0.021$\pm0.011$   & -0.95 & 1.0& 94.2   & \\
RS14 & 9.3587&  0.1414&      5.8     & 0.019$\pm0.008$  & 0.017$\pm0.009$   & -4.74 & 0.3& 54.0   & CONT\\ 
RS15 &10.7036&  0.2149&     12.2     & 0.039$\pm0.011$  & 0.027$\pm0.012$   &  0.57 & 0.2&122.2   & \\
RS16 &10.9596&  0.1187&      8.1     & 0.026$\pm0.009$  & 0.026$\pm0.015$   & -2.01 & 2.7& 74.3   & ZwCl 0041.1-0008\\
RS17 &11.4570& -0.8486&     21.3     & 0.067$\pm0.015$  & 0.068$\pm0.017$   & -0.38 & 1.1& 51.8   & A0095\\
RS18 &11.5751&  0.0167&     10.0     & 0.033$\pm0.010$  & 0.028$\pm0.011$   & -0.85 & 0.1& 83.3   & \\
RS19 &11.6170& -0.2219&      9.5     & 0.031$\pm0.010$  & 0.028$\pm0.011$   &  0.16 & 0.3& 62.9   & $z_{\rm G}=0.26537$\\
RS20 &11.8959& -0.8947&     10.5     & 0.035$\pm0.011$  & 0.020$\pm0.009$   &  0.28 & 1.0& 49.6   & A0101\\
RS21 &13.9519&  0.6488&     29.1     & 0.078$\pm0.014$  & 0.059$\pm0.015$   & -0.30 & 2.7& 67.1   & A0116\\
RS22 &14.0756& -1.1735&    207.3     & 0.621$\pm0.043$  & 1.413$\pm0.071$   &  0.41 &-9.9& 56.1   & A0119, CUT\\
RS23 &15.6415& -0.8327&      7.3     & 0.019$\pm0.007$  & 0.015$\pm0.007$   &  0.14 & 0.0& 74.6   & \\
RS24 &15.6763&  1.1263&     31.9     & 0.077$\pm0.014$  & 0.060$\pm0.014$   &  0.50 & 1.4&128.8   & CONT\\
RS25 &15.9883& -0.8067&      6.6     & 0.016$\pm0.006$  & 0.015$\pm0.015$   & -0.50 & 1.5& 64.7   & \\
RS26 &16.0781& -0.7695&     10.3     & 0.025$\pm0.008$  & 0.012$\pm0.006$   & -2.98 & 0.5& 47.5   & RASSCALS 087\\
RS27 &16.0791& -0.4288&     16.8     & 0.040$\pm0.010$  & 0.023$\pm0.008$   & -0.02 & 1.1&101.6   & CONT\\
RS28 &16.2231&  0.0484&     26.4     & 0.064$\pm0.012$  & 0.062$\pm0.013$   & -2.45 & 2.5&174.5   & ZwCl 0102.4-0012\\
RS29 &16.3765& -0.1699&     14.2     & 0.035$\pm0.009$  & 0.030$\pm0.019$   & -0.60 & 0.9& 64.0   & $z_{\rm G}=0.23133$\\
RS30 &16.5541&  0.8009&      4.1     & 0.122$\pm0.019$  & 0.012$\pm0.006$   &  0.11 & 2.4& 90.0   & \\
RS31 &16.7056&  1.0513&     69.3     & 0.165$\pm0.020$  & 0.149$\pm0.020$   & -0.55 & 0.4& 84.1   & RXCJ 0106.8, CONT\\
RS32 &16.7805& -0.3280&     13.4     & 0.032$\pm0.009$  & 0.032$\pm0.015$   &  0.12 & 0.8&139.2   & $z_{\rm G}=0.26537$\\
RS33 &17.8626&  0.5741&      6.9     & 0.017$\pm0.006$  & 0.015$\pm0.007$   &  0.37 & 0.3&104.2   & $z_{\rm G}=0.36748$, CONT\\
RS34 &18.7749&  0.3502&    188.7     & 0.500$\pm0.036$  & 0.435$\pm0.039$   & -3.35 &24.9& 69.6   & A0168, RXCJ0114.9\\
RS35 &20.5041&  0.3503&     24.5     & 0.049$\pm0.010$  & 0.068$\pm0.020$   &  0.82 & 2.7& 86.9   & A0181, RXCJ0121.9\\
RS36 &21.8232&  0.3523&     27.3     & 0.065$\pm0.012$  & 0.059$\pm0.013$   &  0.29 & 2.4&181.7   & $z_{\rm G}=0.35613$\\
RS37 &22.0578& -0.6777&      8.4     & 0.020$\pm0.007$  & 0.023$\pm0.011$   & -2.47 & 0.2& 66.5   & $z_{\rm G}=0.40152$\\
RS38 &22.6058&  0.4817&      6.1     & 0.014$\pm0.006$  & 0.011$\pm0.006$   & -1.39 & 0.6&107.2   & \\
RS39 &22.9211&  0.5533&     39.1     & 0.095$\pm0.015$  & 0.090$\pm0.018$   & -1.87 & 5.9& 53.1   & A0208, RXC0131\\
RS40 &23.7341& -0.9379&      6.6     & 0.016$\pm0.006$  & 0.016$\pm0.007$   & -1.70 & 0.2& 46.0   & \\
RS41 &24.3671& -0.1861&      7.0     & 0.023$\pm0.009$  & 0.021$\pm0.010$   &  0.62 & 0.0&162.1   & $z_{\rm G}=0.36748$\\
RS42 &26.6947& -0.6782&     39.3     & 0.099$\pm0.016$  & 0.081$\pm0.015$   & -4.09 & 0.1& 34.6   & CONT\\
RS43 &27.2777& -0.4271&     19.6     & 0.049$\pm0.011$  & 0.027$\pm0.015$   &  0.29 & 0.5& 64.0   & $z_{\rm G}=0.33344$\\
RS44 &27.3363& -1.1850&      6.7     & 0.016$\pm0.006$  & 0.023$\pm0.010$   & -0.70 & 0.1&160.9   & $z_{\rm G}=0.29940$\\
RS45 &27.3458& -0.3773&     15.3     & 0.038$\pm0.010$  & 0.011$\pm0.006$   &  0.16 & 0.2& 81.7   & ZwCl 0146.8-0035\\
RS46 &27.8459& -1.0133&      6.2     & 0.015$\pm0.006$  & 0.012$\pm0.006$   & -0.94 & 0.2& 44.3   & \\
RS47 &28.1781&  1.0044&     85.8     & 0.206$\pm0.022$  & 0.209$\pm0.027$   & -1.93 & 3.6&180.5   & A0267, RXCJ0152.7\\
RS48 &28.6185& -0.6531&      9.7     & 0.023$\pm0.007$  & 0.017$\pm0.007$   &  0.14 & 0.0&108.9   & $z_{\rm G}=0.43556$\\
RS49 &29.0463&  1.0481&     23.3     & 0.058$\pm0.012$  & 0.065$\pm0.016$   &  0.14 & 4.2& 96.2   & A0279\\
\hline
\hline
\end{tabular}
\end{center}
\end{table*}

\begin{table*}
{\bf Tab.\,3.} Continued\\
\vspace{-0.1cm}
\begin{center}
\begin{tabular}{rrrrccrrrl}
\hline
\hline
No.  &  R.A. &   DEC. & $N_{\rm ph}$ & $\Lambda$        &  ctr(GCA) & HR & EXT & Richness & Remarks \\
     & 2000.0& 2000.0 &              & $[{\rm s}^{-1}]$ & $[{\rm s}^{-1}]$  & GCA& GCA & SDSS     &         \\ 
\hline		       
RS50 &29.1534&  0.8521&     12.2     & 0.030$\pm0.009$  & 0.023$\pm0.009$   & -0.66 & 0.0&  93.3   & \\
RS51 &29.3443& -0.1020&      8.0     & 0.019$\pm0.007$  & 0.024$\pm0.014$   & -0.08 & 0.4& 174.2   & \\
RS52 &29.4179& -0.1211&     12.7     & 0.032$\pm0.009$  & 0.030$\pm0.017$   &  0.15 & 1.4&  97.7   & \\
RS53 &30.5762& -1.1003&     71.2     & 0.185$\pm0.022$  & 0.190$\pm0.025$   & -3.21 & 4.7&  32.9   & A0295, RXC0202.3\\
RS54 &32.7195& -1.1530&     11.6     & 0.034$\pm0.010$  & 0.027$\pm0.014$   & -0.69 & 1.6&  54.9   & \\
RS55 &33.5548& -0.1419&      7.9     & 0.028$\pm0.010$  & 0.018$\pm0.009$   & -0.91 & 0.0&  48.3   & \\
RS56 &34.6231& -0.4454&      4.8     & 0.027$\pm0.012$  & 0.019$\pm0.012$   &  0.31 & 0.3& 122.0   & \\
RS57 &37.0257& -1.1727&     14.3     & 0.065$\pm0.017$  & 0.053$\pm0.022$   &  0.83 & 0.9&  25.3   & \\
RS58 &38.9997& -1.0232&     10.6     & 0.085$\pm0.026$  & 0.033$\pm0.016$   & -0.77 & 2.4&  71.1   & $t_{\rm eff}=124.8$\,s\\
RS59 &41.4703& -0.6960&      5.7     & 0.067$\pm0.028$  & 0.033$\pm0.021$   &  0.37 & 0.4& 118.7   & $t_{\rm eff}= 85.8$\,s, A0381\\
RS60 &42.0287& -0.6403&      6.8     & 0.055$\pm0.021$  & 0.045$\pm0.020$   & -1.33 & 0.5&  39.3   & GGrp $z=0.02056$\\
RS61 &42.3519&  1.0279&     11.5     & 0.060$\pm0.018$  & 0.052$\pm0.020$   & -0.57 & 0.6&  99.5   & $z_{\rm G}=0.26537$\\
RS62 &46.0947&  1.0847&      6.1     & 0.024$\pm0.010$  & 0.016$\pm0.008$   &  0.12 & 1.5&  82.4   & A0411, RXCJ0304\\
RS63 &46.1463& -0.8985&      9.5     & 0.044$\pm0.014$  & 0.034$\pm0.014$   & -0.37 & 0.4&  91.2   & CONT\\
RS64 &46.5799& -0.2233&     15.3     & 0.065$\pm0.017$  & 0.021$\pm0.010$   &  0.16 & 1.2&  37.2   & A0412\\
RS65 &47.9288&  1.1583&      7.5     & 0.017$\pm0.006$  & 0.013$\pm0.006$   & -0.68 & 0.0&  85.5   & $z_{\rm G}=0.19729$\\
RS66 &50.1187&  0.5485&     18.1     & 0.054$\pm0.013$  & 0.042$\pm0.015$   & -0.70 & 0.3&  89.8   & \\
RS67 &52.0082&  0.4007&      7.6     & 0.022$\pm0.008$  & 0.014$\pm0.006$   & -1.47 & 0.0& 150.2   & \\
RS68 &52.1520& -0.0535&     19.5     & 0.055$\pm0.012$  & 0.046$\pm0.013$   &  0.29 & 0.6&  68.1   & $z_{\rm G}=0.26537$, CONT\\
RS69 &52.6521& -0.5483&     16.3     & 0.039$\pm0.010$  & 0.019$\pm0.007$   & -0.49 & 0.9&  54.5   & \\
RS70 &53.1336&  0.3992&      7.4     & 0.014$\pm0.005$  & 0.011$\pm0.006$   & -1.49 & 0.0& 168.5   & $z_{\rm G}=0.39017$\\
RS71 &53.5782& -1.1857&     41.9     & 0.080$\pm0.012$  & 0.107$\pm0.019$   &  0.24 & 3.2&  86.6   & RXCJ0334.0\\
RS72 &53.6314& -0.7342&     21.6     & 0.040$\pm0.009$  & 0.037$\pm0.010$   & -0.26 & 2.0& 121.7   & $z_{\rm G}=0.41286$\\
RS73 &53.9795& -0.6320&     12.3     & 0.024$\pm0.007$  & 0.011$\pm0.007$   &  0.02 & 1.3&  71.3   & $z_{\rm G}=0.26537$\\
RS74 &55.6623& -0.2708&     11.0     & 0.030$\pm0.009$  & 0.016$\pm0.016$   &  0.11 & 0.9& 190.7   & $z_{\rm G}=0.35613$\\
RS75 &56.3280&  1.0026&     17.0     & 0.052$\pm0.013$  & 0.013$\pm0.007$   &  0.07 & 3.4&  57.3   & ZwCl 0342.8+0052\\
\hline
\hline
\end{tabular}
\end{center}
\end{table*}

\clearpage

\end{document}